\documentclass[longauth]{aa}
\usepackage{graphicx}
\usepackage{txfonts}
\usepackage{bbding}
\usepackage{float}
\usepackage{upgreek}
\usepackage{stmaryrd}
\usepackage{supertabular}
\usepackage[normalem]{ulem}
\usepackage{hyperref}
\hypersetup{
    colorlinks=true,
    citecolor=blue,
    linkcolor=blue,
    filecolor=blue,
    urlcolor=blue
    }
\newcommand{\Rjup}{\ensuremath{\mathrm{R_{Jup}}}\xspace}
\newcommand{\MJup}{\ensuremath{\mathrm{M_{Jup}}}\xspace}
\newcommand{\Teff}{\ensuremath{\mathrm{T_{eff}}}\xspace}
\newcommand{\Mjup}{\MJup}

\newcommand{\Rlambda}{\ensuremath{\mathrm{R_\lambda}}}
\newcommand{\fsed}{\ensuremath{\mathrm{f_{sed}}}}
\newcommand{\Kzz}{\ensuremath{\mathrm{K_{zz}}}}
\newcommand{\Lclassic}{\mathcal{L}^{\mathrm{classic}}}
\newcommand{\Lgp}{\mathcal{L}^{\mathrm{GP}}}

\begin{document}

\title{Panchromatic View of the Frigid Jovian Exoplanet COCONUTS-2~b}

\author{
        Matthieu Ravet\inst{\ref{lagrange}, \ref{IPAG}, \ref{MPIA}}
        \and
        Mickaël Bonnefoy\inst{\ref{IPAG}}
        \and
        Gaël Chauvin\inst{\ref{MPIA}}
        \and
        Zhoujian Zhang\inst{\ref{SC}}
        \and
        Jacqueline K. Faherty\inst{\ref{DA}}
        \and
        Maël Voyer\inst{\ref{CEA}}
        \and
        Mark W. Phillips\inst{\ref{EDIN}}
        \and
        Pascal Tremblin\inst{\ref{CEA}}
        \and
        Rocio Kiman\inst{\ref{PAS}}
        \and
        Jessica Copeland\inst{\ref{HERT}}
        \and
        James J. Mang\inst{\ref{TEX}}
        \and
        Caroline V. Morley\inst{\ref{TEX}}
        \and
        Helena Kühnle\inst{\ref{ETH}}
        \and
        Benjamin Charnay\inst{\ref{LIRA}}
        \and
        Sam de Regt\inst{\ref{Leiden}}
        \and
        Paul Mollière\inst{\ref{MPIA}}
        \and
        Simon Petrus\inst{\ref{NASA}, \ref{IEAFICUD}, \ref{YEMS}}
        \and
        Allan Denis\inst{\ref{LAM}}
        \and
        Alice Radcliffe\inst{\ref{LIRA}}
        \and
        Paulina Palma-Bifani\inst{\ref{LIRA}}
        \and
        Arthur Vigan\inst{\ref{LAM}}
        \and
        Mathilde Mâlin\inst{\ref{JHU}, \ref{STSCI}, \ref{LIRA}}
        \and
        Gabriel-Dominique Marleau\inst{\ref{FFP}, \ref{PIuB}, \ref{MPIA}}
        \and
        Elena Manjavacas\inst{\ref{JHU}, \ref{AURA}}
        \and
        Kevin Hoy\inst{\ref{IEAFICUD}, \ref{YEMS}, \ref{ESO}}
        \and
        Elisabeth C. Matthews\inst{\ref{MPIA}}
        \and
        Thomas K. Henning\inst{\ref{MPIA}}
}

\institute{            
            Laboratoire J.-L. Lagrange, Université Côte d'Azur, Observatoire de la Côte d'Azur, CNRS, 06304 Nice, France\label{lagrange}
            \and
            IPAG, Université Grenoble-Alpes, CNRS, F-38000 Grenoble, France\label{IPAG}
            \and
            Max-Planck-Institut für Astronomie, Königstuhl 17, 69117 Heidelberg, Germany\label{MPIA}
            \and
            Department of Physics \& Astronomy, University of Rochester, Rochester, NY 14627, USA\label{SC}
            \and
            Department of Astrophysics, American Museum of Natural History, New York, NY 10024, USA\label{DA}
            \and
            Université Paris Cité, Université Paris-Saclay, CEA, CNRS, AIM, F-91191 Gif-sur-Yvette, France\label{CEA}
            \and
            Institute for Astronomy, University of Edinburgh, Royal Observatory, Blackford Hill, Edinburgh, EH9 3HJ, UK\label{EDIN}
            \and
            Department of Astronomy, California Institute of Technology, Pasadena, CA 91125, USA\label{PAS}
            \and
            Department of Physics, Astronomy and Mathematics, University of Hertfordshire, Hatfield, UK\label{HERT}
            \and
            Department of Astronomy, University of Texas at Austin, Austin, TX 78712, USA\label{TEX}
            \and
            Institute of Particle Physics and Astrophysics, ETH Zürich, Wolfgang-Pauli-Str 27, 8049 Zürich Switzerland\label{ETH}
            \and
            LIRA, Observatoire de Paris, Univ. PSL, CNRS, Sorbonne Universit´e, Univ. Paris Diderot, Sorbonne Paris Cit´e, 5 place Jules Janssen, 92195 Meudon, France\label{LIRA}
            \and
            Leiden Observatory, Leiden University, P.O. Box 9513, 2300 RA, Leiden, The Netherlands\label{Leiden}
            \and
            NASA-Goddard Space Flight Center, Greenbelt, MD 20771, USA\label{NASA}
            \and
            Instituto de Estudios Astrof\'isicos, Facultad de Ingenier\'ia y Ciencias, Universidad Diego Portales, Av.\ Ej\'ercito Libertador 441, Santiago, Chile\label{IEAFICUD}
            \and
            Millennium Nucleus on Young Exoplanets and their Moons (YEMS), Santiago, Chile\label{YEMS}
            \and
            Aix Marseille Univ, CNRS, CNES, LAM, Marseille, France\label{LAM}
            \and
            Department of Physics \& Astronomy, Johns Hopkins University, Baltimore, MD, 21218, USA\label{JHU}
            \and
            Space Telescope Science Institute, 3700 San Martin Drive, Baltimore, MD 21218, USA\label{STSCI}
            \and
            Fakultät für Physik, Universität Duisburg-Essen, Lotharstraße 1, 47057 Duisburg, Germany\label{FFP}
            \and
            Physikalisches Institut, Universität Bern, Gesellschaftsstr. 6, CH-3012 Bern, Switzerland\label{PIuB}
            \and
            AURA for the European Space Agency (ESA), ESA Ofcice, Space Telescope Science Institute, 3700 San Martin Drive, Baltimore, MD, 21218 USA\label{AURA}
            \and
            European Southern Observatory, Alonso de C\'ordova 3107, Casilla 19, Santiago 19001, Chile\label{ESO}
}

\date{Received XXXX; accepted YYYY}

   \date{Received XXXX; accepted YYYY}
 
  \abstract
   {Cold exo-Jovian planets are beginning to be imaged and characterised using the James Webb Space Telescope (JWST) instruments. These observations often reveal new molecular species (CO$_2$, NH$_3$, PH$_3$), challenge atmospheric models and raise questions about the formation pathways and evolution of these objects.}
   {We revisit the atmosphere of the cold (\Teff~$=483^{+44}_{-53}$~K), mature ($414\pm23$~Myr) and large separation ($>5000$~au) Jovian exoplanet COCONUTS-2~b (WISEPA J075108.79-763449.6), adding new spectral information beyond 5~$\mu$m and combining them with existing spectrophotometry to consolidate the constraints on the object properties and identify disagreements from self-consistent atmospheric models.}
   {We use a high signal-to-noise MIRI-LRS spectrum (5.45--11~$\mu$m, \Rlambda~$\sim100$) of COCONUTS-2~b revealing prominent molecular features of H$_2$O, CH$_4$ and NH$_3$. This dataset is combined with spectra from Gemini/FLAMINGOS-2 and JWST/NIRSpec (G395H), as well as photometry from WISE and Spitzer, resulting in almost continuous wavelength coverage from 1 to 15~$\mu$m. We analyze the data using five grids of self-consistent atmospheric models, spanning a wide range of \Teff, log(g), and [M/H]. We also investigate the use of Gaussian Processes to account for correlated noise either caused by the spectrograph or by systematic departures of models in the inversion framework.}
   {All models manage to fit the overall combined observations but predict fainter flux in Y- and N-bands. Classical model comparison suggests that the ATMO2020++ synthetic specra (with and without PH$_3$) are statistically  preferred. However, when accounting for correlated noise using Gaussian processes, Sonora Elf Owl models become favoured; although they still provide a comparatively poor fit to the data with bulk properties inconsistent with cooling models predictions. Fitting for the correlated noise of the three spectroscopic instruments, ATMO2020++ models yields constraints consistent with previous studies and evolutionary models predictions: \Teff~$=496^{+5}_{-3}$~K, log(g)~$=4.30^{+0.04}_{-0.02}$~dex, [M/H]~$=-0.02^{+0.03}_{-0.02}$~dex, and R~$=1.03^{+0.01}_{-0.02}$~\Rjup. The extended wavelength coverage provided by MIRI (accounting for 41\% of the bolometric flux) completes the SED, yielding a precise luminosity estimation of log(L/L$_{\odot}$)~$=-6.166\pm0.002$~dex. Combined with a previous estimate of the system age ($414\pm23$~Myr), cooling models predict a mass of M~$=7.3\pm0.3$~\Mjup.}
   {The preferred models suggest a metallicity consistent with that of the primary, potentially supporting a binary-like formation scenario. Remaining discrepancies across spectral bands and between model grids suggest incomplete chemistry modeling and highlight the need for improved treatments of alkali condensation and diabatic processes for models at these low effective temperatures}

   \keywords{Techniques: high angular resolution, spectroscopic; Methods: data analysis, observational, statistical; Planets and satellites: atmospheres, gaseous planets}

   \maketitle

\section{Introduction}

Although only a few planetary-mass companions have been detected and characterized using direct imaging\footnote{\url{https://exoplanetarchive.ipac.caltech.edu/}} (approximately 80), this sample includes diverse objects with effective temperatures (\Teff) ranging from 270 to 2700~K and orbital separations spanning from a few tens to thousands of au. The majority of these companions are young (<~300~Myr) and the observed diversity of properties (e.g., log(L/L$_{\odot}$), \Teff, log(g), C/O, [M/H], $e$) is a product of each objects formation and evolution history \citep[e.g.,][]{Spiegel2012, Allers2013, Ruffio2026}.

\begin{table}[t!]
    \setlength{\tabcolsep}{6pt}
    \renewcommand{\arraystretch}{1.2} 
    \centering
    \caption{\centering \small{Literature physical properties of the COCONUTS-2 system}}
    \begin{tabular}{lll}
    \hline
    \hline
    \multicolumn{2}{c}{COCONUTS-2~A} & Refs. \\
    \hline
    Spectral type & M3 & (a) \\
    \Teff (K) & 3406$\pm$69 & (c) \\
    log(g) & 4.83$\pm$0.03 & (h) \\
    \text{[M/H]} & 0.00$\pm$0.08 & (d) \\
    Distance (pc) & 10.888$\pm$0.002 & (f) \\
    Radial velocity (km.s$^{-1}$) & 1.19$\pm$0.61 & (e) \\
    Age (Myr) & 414$\pm$23 & (j) \\
    Mass (M$_{\odot}$) & 0.37$\pm$0.011 & (h) \\
    Radius (R$_{\odot}$) & 0.388$\pm$0.011 & (h) \\
    \hline
    \\
    \multicolumn{2}{c}{COCONUTS-2~b} & \\
    \hline
    Spectral type & T9 & (b) \\
    \Teff (K) & 483$^{+44}_{-53}$ & (i) \\
    log(g) & 4.19$^{+0.18}_{-0.13}$ & (i) \\
    Semimajor axis (au) & 7506$^{+5205}_{-2060}$ & (h+)\\
    Period (Myr) & 1.1$^{+1.3}_{-0.4}$ & (h+) \\
    Age (Myr) & 414$\pm$23 & (j) \\
    Mass (\Mjup) & 8$\pm$2 & (i) \\
    Radius (\Rjup) & 1.11$^{+0.03}_{-0.04}$ & (i) \\
    \hline
    \end{tabular}
    \tablefoot{\small{ \tablefoottext{a}{\small\cite{torres_search_2006}}, \tablefoottext{b}{\small\cite{kirkpatrick_first_2011}}, \tablefoottext{c}{\small\cite{gaidos_trumpeting_2014}}, \tablefoottext{d}{\small\cite{hojjatpanah_catalog_2019}}, \tablefoottext{e}{\small\cite{schneider_acronym_2019}}, \tablefoottext{f}{\small\cite{gaia_collaboration_gaia_2021}}, \tablefoottext{g}{\small\cite{bailer-jones_estimating_2021}}, \tablefoottext{h}{\small\cite{zhang_second_2021}}, \tablefoottext{i}{\small\cite{zhang_disequilibrium_2024}} and \tablefoottext{j}{\small\cite{kiman_diversity_2025}}. The semi-major axis and orbital period were estimated using orbital predictions based on the eccentricity distributions from \cite{dupuy_distribution_2011} and \cite{zhang_second_2021} measured projected separation of 594".}}
    \label{table_sys}
\end{table}

The COCONUTS-2 (L 34-26 and WISEPA J075108.79-763449.6) system stands out as an outlier in the field of direct imaging. It consists of a cold $\sim$8~\Mjup \citep{zhang_disequilibrium_2024}, T9 \citep{kirkpatrick_first_2011} companion orbiting a M3 star with an estimated age of $414\pm23$~Myr \citep{kiman_diversity_2025}, providing a unique opportunity to study the atmosphere of a mature super-Jupiter. Originally thought to be two separate free-floating objects, \cite{zhang_cool_2020, zhang_second_2021} demonstrated, through proper motion and parallax measurements, that COCONUTS-2~A and b are gravitationally bound. They estimated the semi-major axis of COCONUTS-2~b to be 7506$^{+5205}_{-2060}$~au and its effective temperature to be \Teff~=~434$\pm$9~K, making it the second coldest and largest separation directly imaged exoplanet discovered to date. The large separation poses a challenge for planet formation models. COCONUTS-2b may have formed within the circumstellar disk of A, either through core accretion \citep{pollack_formation_1996, ida_toward_2008, alibert_models_2005, alibert_theoretical_2013, benz_planet_2014} or disk instability \citep{kuiper_origin_1951, cameron_physics_1978, boss_giant_1997}, followed by outward migration due to dynamical interactions \citep{vorobyov_formation_2013}. Alternatively, COCONUTS-2b could have formed via gravitational instability \citep{padoan_stellar_2002} in a collapsing molecular cloud, similar to a stellar binary \citep{zuckerman_minimum_2009, kirkpatrick_first_2011}. A less likely scenario is a capture during a flyby, as suggested by \cite{marocco_thirteen_2024}. The recent extensive study from \cite{zhang_disequilibrium_2024} using the combination of WISE and Spitzer photometry with Gemini/FLAMINGO-2 spectroscopy (0.98--2.51~$\mu$m, \Rlambda~$\sim200-1200$) alongside multiple atmospheric grids revealed that COCONUTS-2~b likely has a sub- or near-stellar C/O ratio and metallicity ([M/H]). This may suggests that it accreted oxygen-rich and carbon-depleted gas from within ice lines \citep{line_solar_2021, august_confirmation_2023} and was not influenced by core erosion processes nor planetesimal bombardment \citep[e.g.,][]{alibert_theoretical_2013, miller_heavy-element_2011, thorngren_connecting_2019, madhusudhan_exoplanetary_2019, schneider_how_2021, zhang_elemental_2023}. This could indicate a planet-like formation followed by outward migration. However, the spread of values in their retrieved metallicities accross models (ranging from -0.4 to 0.1) remains fairly consistent with the metallicity of COCONUTS-2~A ($0.00\pm0.08$, \citealp{hojjatpanah_catalog_2019}) and therefore cannot be used to rule out a scenario of a binary-like formation, as the metallicities of the host star and the companion coincide.

We present in this study new observations from JWST/MIRI-LRS that we analyse jointly with existing spectroscopic data to consolidate the estimate of the physical properties of COCONUTS-2~b. The combination of the very high S/N of these data and the low effective temperature of COCONUTS-2~b presents a challenge for classical forward modeling approaches \citep{petrus_jwst_2024}. These temperature and age regimes remain largely unexplored, and self-consistent models may suffer from systematic deviations, either due to missing or incomplete physics (e.g., disequilibrium chemistry, inhomogeneous atmospheres), or due to limitations in parameter space exploration. Our analysis method is presented in Sect.~\ref{sec2}, and the results of our analysis are given in Sect.~\ref{sec3}. We discuss the implications of our results in Sect.~\ref{sec4}, and present our conclusions in Sect.~\ref{sec5} Additional material is provided in the Appendices.

\section{Methods}\label{sec2}

\subsection{Observation and data reduction}

We use both archival and newly obtained spectro-photometric data on the target. This section describes each of them and in Table~\ref{table_archival} we provide a concise summary of all of them.

\paragraph{Photometry:} COCONUTS-2~b was observed with both Spitzer and WISE as part of a survey of 184 late-T and Y dwarfs presented in \cite{kirkpatrick_preliminary_2019}. We used the Spitzer [3.6] and [4.5] points, as well as the WISE W1 ($3.32\pm0.33$~$\mu$m) and W2 ($4.56\pm0.52$~$\mu$m). The W3 and W4 bands were excluded, as their bandpasses are not fully covered by the Sonora Elf Owl grid, with W4 lying entirely outside the modeled wavelength range. We note that this has little impact on the retrieved parameters for the other grids.

\paragraph{Gemini/FLAMINGOS-2:} FLAMINGOS-2 is a low/medium-resolution spectrograph located on the 8.1-m Gemini-South telescope. We use archival FLAMINGOS-2 observations of COCONUTS-2~b obtained as part of programs GS-2021B-FT-111 and GS-2021B-FT-204 and first presented in \citealp{zhang_disequilibrium_2024}. The instrument provides simultaneous coverage of the Y-, J-, H- and K-bands (1.01--2.50~$\mu$m). COCONUTS-2~b was observed on the 16 and 19 December 2021 with the 2-pixel-wide slit (0.36"$\times$263") and an ABBA nodding pattern, resulting in an average spectral resolution of  \Rlambda~$\sim900$ \citep{zhang_disequilibrium_2024}. Because COCONUTS-2~A and b are widely separated, the adaptative optics (AO) system was not needed. To extract the final spectrum, a standard data reduction was applied using the \texttt{\textit{Gemini-IRAF}}\footnote{\url{https://gemini-iraf-vm-tutorial.readthedocs.io/en/stable/}} package.

\paragraph{JWST/NIRSpec:} As part of the "Explaining the Diversity of Cold Worlds" survey (Program ID: 2124, PI J. Faherty, \href{https://doi.org/10.17909/rxm9-qd05}{DOI}), COCONUTS-2~b was observed with JWST/NIRSpec. Observations took place on the 9 July 2023 for a total integration time of 2777.706~s and where reduced and presented in \citealp{kiman_diversity_2025}. They used the G395H grating with the F290LP filter element resulting in an average spectral resolution of \Rlambda~$\sim3000$ and a wavelength coverage of 2.88--5.14~$\mu$m. We convolved the this spectra to a lower resolution of \Rlambda~$\sim1500$ in our analyis to match the sampling of the ATMO2020++ grids.

\paragraph{JWST/MIRI-LRS:} MIRI-LRS on the JWST is a low resolution spectrograph operating in the mid infrared. It was used to observe COCONUTS-2~b on the 29 March 2024 (Program ID: 3514, PI M. Bonnefoy, \href{https://doi.org/10.17909/p6kh-4h93}{DOI}) for a total integration time of 3335.598~s. The observation strategy was based on using a two-dither pattern with two integrations of 300 groups per exposure. The data were reduced using a combination of the official JWST STScI pipeline\footnote{\url{https://zenodo.org/records/17515973}}, supplemented with customized functions described in \cite{Voyer_wd0806_2025} section 2.2. As noted in other studies \citep{2024ApJ...977L..32X, Voyer_wd0806_2025} the retrievals showed a significant wavelength-dependent offset between line positions of the data and the theoretical models. To correct these offsets we include a $3^{rd}$ order polynomia for the wavelength correction in the retrieval. The coefficients of this correction are retrieved once and then fixed for subsequent analysis folloing the methods in \cite{2024ApJ...977L..32X, Voyer_wd0806_2025}. The wavelength solution from Table 1\footnote{\url{https://content.cld.iop.org/journals/2041-8205/982/2/L38/revision2/apjladbd46t1_mrt.txt}} of \cite{Voyer_wd0806_2025} was used for the pixel-to-wavelength calibration. Beyond $12.5~\mu$m, the extracted spectrum is dominated by background emission. With only two integrations per exposure, the background cannot be reliably averaged out, and the resulting data quality is insufficient for scientific analysis. We therefore exclude these longer wavelengths from the analysis.

\begingroup

\setlength{\tabcolsep}{6pt} 
\renewcommand{\arraystretch}{1.5} 
\begin{table*}[h!]
    \centering
    \caption{\centering \small{Summary of the new and archival data used.}}
    \begin{tabular}{lllllll}
    \hline
    \hline
    Instrument & Type & Spectral coverage & Spectral resolution & S/N & Refs. \\
    \hline
    WISE & (W1) L-band photometry & 3.32$\pm$0.33 $\mu$m & <30 & $\sim$69 & (a) \\
    & (W2) M-band photometry & 4.56$\pm$0.52 $\mu$m & <30 & $\sim$167 & (a) \\
    \hline
    Spitzer & (c1) L-band photometry & 3.51$\pm$0.34 $\mu$m & <30 & $\sim$69 & (a) \\
    & (c2) M-band photometry & 4.44$\pm$0.43 $\mu$m & <30 & $\sim$125 & (a) \\
    \hline
    Gemini/FLAMINGOS-2 & YJHK-band spectroscopy & 1.01--2.50 $\mu$m & 200--1200 & 2--10 & (b) \\
    \hline
    JWST/NIRSpec & G395H/F290LP LM-band spectroscopy & 2.88--5.14 $\mu$m & 2000--3500 & 10--60 & (c) \\
    \hline
    JWST/MIRI-LRS & P750L N-band spectroscopy & 5.00--12.5 $\mu$m & 30--200 & 50--700 & (d) \\
    \hline
    \end{tabular}
    \tablefoot{\small{The first column contains the name of the instrument, the second the type (spectroscopy or photometry), the third the wavelength coverage, the forth the (effective) spectral resolution \Rlambda, the fifth the signal-to-noise ratio (S/N). References:  \tablefoottext{a}{\small\cite{kirkpatrick_preliminary_2019}}, \tablefoottext{b}{\small\cite{zhang_disequilibrium_2024}}, \tablefoottext{c}{\small\cite{kiman_diversity_2025}} \tablefoottext{d}{\small(this work)}.}}
    \label{table_archival}
\end{table*}

\endgroup

\subsection{Framework}\label{sec2.2}

To infer the atmospheric properties of COCONUTS-2~b, we used a custom version of the Forward Modeling tool for Spectral Analysis, \texttt{\textit{ForMoSA}}\footnote{\url{https://formosa.readthedocs.io/en/latest/}} \citep{ravet_mutli_2025}. It relies on the nested-sampling \citep{nested_skilling_2004} approach using the python package \texttt{\textit{PyMultinest}}\footnote{\url{https://johannesbuchner.github.io/PyMultiNest/}} \citep{buchner_pmn_2014}. Each inversion utilise 1000 live points, adaptive sampling efficiency mode, an evidence tolerance of 0.5 and uniform priors where used for the atmospheric parameters. We also fit for the planet radius R constrained by the flux dilution factor $\left(\frac{\text{R}}{d}\right)^2$ using a fixed distance of 10.888~($\pm0.002$)~pc (see Table~\ref{table_sys}). Nested sampling allows the direct computation of the log-Bayesian evidence, $\ln \mathcal{Z}$, enabling robust, quantitative comparison between models via the log-Bayes factor, $\ln \mathcal{B}_{1,2}=\ln \mathcal{Z}_1-\ln \mathcal{Z}_2$. This metric inherently accounts for both model fit and complexity (i.e., number of free parameters). To facilitate intuitive interpretation, the Bayes factor is often approximately translated into a "sigma" significance level, representing the strength of preference for one model over another \citep{sellke_calibration_2001, trotta_bayes_2008, benneke_how_2013}. However, as recently pointed out by \citet{kipping_exoplaneteers_2025}, such a conversion may only provide an upper limit on the "sigma" values. The true values are generally lower, which can lead to an overestimation of the significance of model preferences. We therefore chose to rely on the log-Bayesian evidence as a more conservative proxy for model significance. Importantly, the trend we find is also consistent with, Akaike Information Criterion (AIC, \citealp{Akaike1974}), the Bayesian Information Criterion (BIC, e.g., \citealp{molliere_model_selection_2025}) and Simplified Bayesian Predictive Information Criterion (BPICS, \citealp{Ando2011, Thorngren2025})  across all tested models (see Table.~\ref{bayes_classic}). For the rest of the study, we define these quantities relative to the favored model:
\begin{equation}
\begin{split}
    & \ln \mathcal{B} = \ln \mathcal{Z}_{max} - \ln \mathcal{Z}, \\
    & \Delta\text{AIC} = \text{AIC} - \text{AIC}_{min}, \\
    & \Delta\text{BIC} = \text{BIC} - \text{BIC}_{min}, \\
    & \Delta\text{BPICS} = \text{BPICS} - \text{BPICS}_{min},
\end{split}
\label{equ1}
\end{equation}
where subscripts $max$ and $min$ denote the maximum and minimum values across all models. With this convention, the preferred model has $\ln \mathcal{B}=0$, while all other models have $\ln \mathcal{B}>0$.

\subsection{Likelihood mapping}

Since using only the injected error bars coming from the data-reduction pipelines likely underestimates the total uncertainty, we introduce in \texttt{\textit{ForMoSA}} the following modified log-likelihood function:

\begin{equation}
\begin{split}
    & \ln \Lclassic = \sum_i \ln \Lclassic_i \\
    & \ln \Lclassic_i = -\frac{N_i}{2}\ln\left( \frac{\chi_i^2}{N_i}\right) - \frac{1}{2}\ln(|\vec{\sigma_i}|) -\frac{N_i}{2}\ln(2\pi) -\frac{N_i}{2}\\
    &\text{with} \quad \chi_i^2 = \sum \left( \frac{\vec{d_i}-\vec{m_i}}{\vec{\sigma_i}} \right)^2,
\end{split}
\label{equ1}
\end{equation}
where the index $i$ takes values from 1 to 4 and corresponds to the three spectroscopic observations and four photometric points. Here, $\vec{d_i}$, $\vec{\sigma_i}$, and $\vec{m_i}$ refer to the data, error, and model vectors respectively, each of length $N_i$. This likelihood formulation proposed by \citet{ruffio_radial_2019} is obtained by marginalizing the standard log-likelihood over a noise scaling factor $\vec{\sigma_s} = s \vec{\sigma}$. The maximum likelihood estimator for this scaling is:

\begin{equation}
    \hat{s_i} = \frac{\chi_i^2}{N_i},
\end{equation}

This approach partially compensates for model–data mismatches by inflating the error bars, thus propagating them into the posterior distributions. Other approaches exist in the literature, including alternative multiplicative or additive prescriptions \citep[e.g.,][]{piette_considerations_2020, zhang_disequilibrium_2024}. We here adopt a marginalization over a simple multiplicative factor as a computationally efficient choice that does not assume a particular functional form of the inflation. In a forward-modeling framework, the quality of the fit is primarily limited by the fidelity of the atmospheric models rather than by the formal observational uncertainties \citep{petrus_xshyne_2025}, making the inclusion of these noise scaling parameters a pragmatic way to account for residual model deficiencies without biasing the posteriors. However, they entirely neglects potential spectral correlations introduced by the spectrograph or by systmematic deviations between data and model spread over large bandwidths. To address this limitation, we employ Gaussian Processes (GP) to model correlated noise between neighbouring pixels. Building upon previous work \citep{czekala_constructing_2015, wang_kecknirc2_2020, de_regt_eso_2024, de_regt_eso_2025, rotman_enabling_2025}, we define the covariance matrix as:

\begin{equation}
    C_{i;\;m,n} = \sigma_{i;\;n}^2\;\delta_{m,n} + a_{i}^2\tilde{\sigma_{i}}^2 \exp\left[-\frac{\Delta \lambda_{m,n}^2}{2\ell_{i}^2}\right],
\end{equation}
where each $C_{i;\;m,n}$ represents the total covariance between pixels $m$ and $n$ in the $i$-th spectrum. The matrix comprises two components: diagonal terms $\sigma_{i;\;n}^2\;\delta_{m,n}$ accounting for uncorrelated noise, and Radial Basis Function (RBF) kernel terms modeling spectral correlations. The RBF kernel is parameterized by an amplitude $a_{i}$, a length scale $\ell_{i}$ and is normalized by the squared median uncertainty $\tilde{\sigma_{i}}^2$. $\Delta \lambda_{m,n}$ denotes the wavelength separation between data points $m$ and $n$. Injecting this covariance matrix $\vec{C_i}$ into the standard log-likelihood, we obtain:

\begin{equation}
\begin{split}
    & \ln \Lgp = \sum_i \ln \Lgp_i \\
    & \ln \Lgp_i = -\frac{\chi_i^2}{2} - \frac{1}{2} \ln (|\vec{C_i}|) -\frac{N_i}{2} \ln (2\pi)\\\\
    &\text{with} \quad \chi_i^2 = (\vec{d_i}-\vec{m_i})^T \vec{C_i}^{-1} (\vec{d_i}-\vec{m_i}),
\end{split}
\label{equ2}
\end{equation}

This likelihood is very similar to the one presented in \citet{de_regt_eso_2025}, except that we do not fit for a noise scaling factor. This choice is motivated by the fact that including a noise scaling term introduces strong biases in the model-data residuals for MIRI-LRS (see bottom panel of Fig.~\ref{fig1_comp_models}, bottom panels of Fig.~\ref{fig_GPvsclassic_MIRI} and discussion in Sect.~\ref{sec4.2}). Equations \ref{equ1} and \ref{equ2} assume that each dataset is independent, a reasonable assumption given that each observation was taken with a different instrument at a different time. Each log($a_{i}$) and log($\ell_{i}$) are retrieved as free parameters during the inversion.

\section{Results}\label{sec3}

We divided our analysis into three parts. First, we performed an exploration using different atmospheric models evaluated with a simple noise scaling (equ. \ref{equ1}). Then, we carried out a more detailed investigation using GP-aided forward modeling (equ. \ref{equ2}), focusing on the best-fitting models identified in the first stage. Finally, we used the preferred solutions to re-derive the bolometric luminosity using the full spectral energy distribution (SED) and, in combination with evolutionary models, infer the corresponding bulk properties. The aim of this strategy is to mitigate some of the limitations of current forward modelling approaches (such as model systematics and correlated noise) to derive more robust constraints on the atmospheric parameters of COCONUTS-2~b.

Similar to \cite{zhang_disequilibrium_2024}, we mask five regions of the FLAMINGOS-2 observation and performed the fit on the following ranges: 1.05--1.12~$\mu$m,
1.18--1.33~$\mu$m, and 1.52--1.66~$\mu$m (see Fig.~\ref{fig1_comp_models}). Fluxes from wavelengths shorter than 1.05~$\mu$m were masked to avoid the area with low S/N near the edge of the detector and valleys among the Y/J/H/K peaks due to the residuals from telluric correction and the relatively large flux uncertainties in the spectrum. We also ignore any variability and do not adjust for relative flux offsets between instruments, as no variability has been reported for COCONUTS-2~b.

\subsection{Model comparison}\label{sec3.1}

\begin{figure*}[t]
\centering
    \includegraphics[scale=0.55]{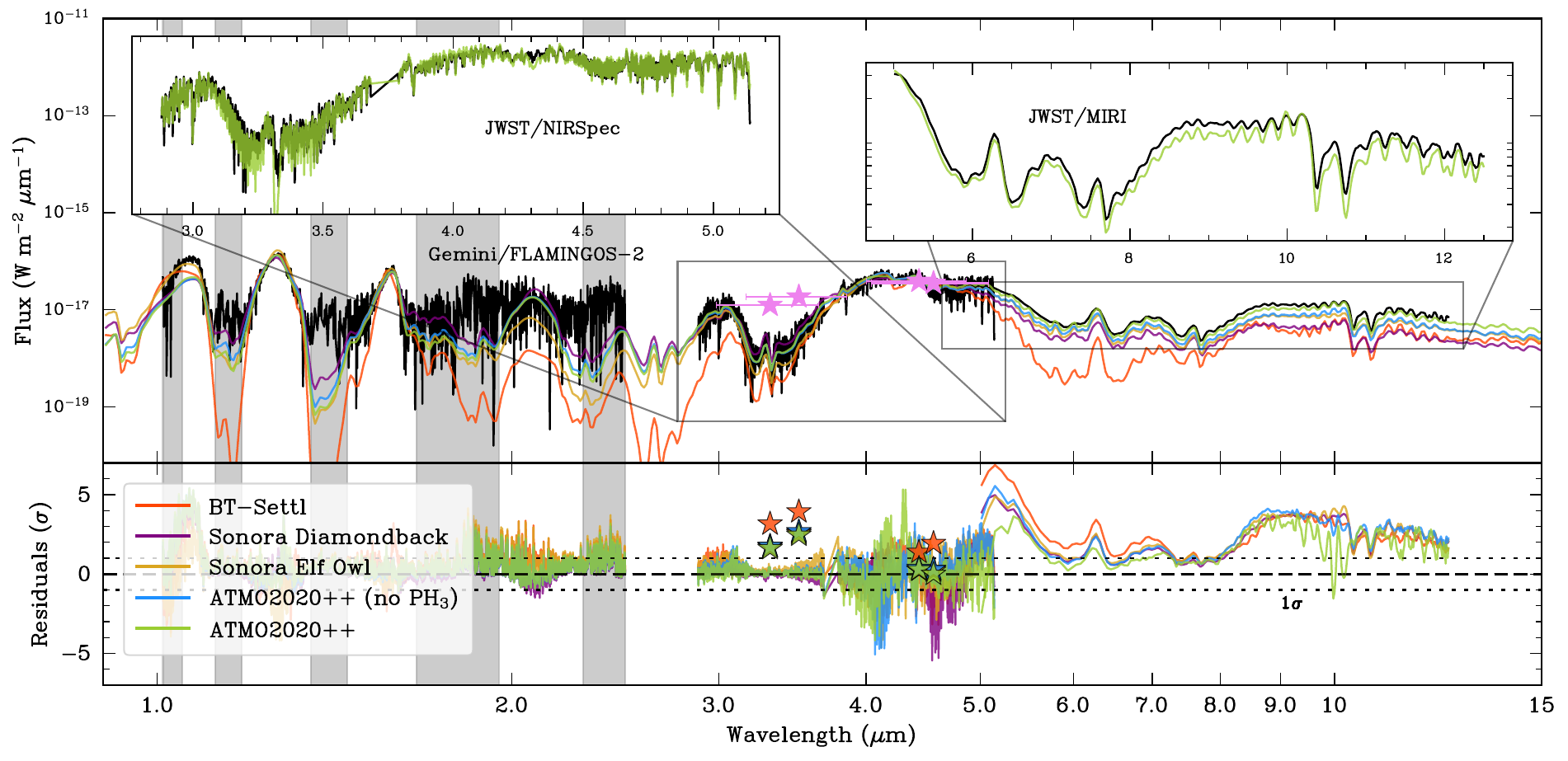}
    \caption{\small{\textit{Top panel:} Forward modeling results of the combined COCONUTS-2~b observations. The black solid lines represent the spectroscopic data (from left to right: Gemini/FLAMINGOS-2, JWST/NIRSpec, and JWST/MIRI-LRS) while colored lines represents each \Rlambda~$\sim100$ model ("classic"). Pink and white stars indicate the photometric observations (WISE and Spitzer). Grey-shaded regions indicate the masked wavelengths during the spectral inversion. \textit{Bottom panel:} Residuals for each fit. Dotted lines represents the $\pm\sigma$ (68\%) confidence interval}}
    \label{fig1_comp_models}
\end{figure*}

\begin{figure*}[t]
\centering
    \includegraphics[scale=0.75]{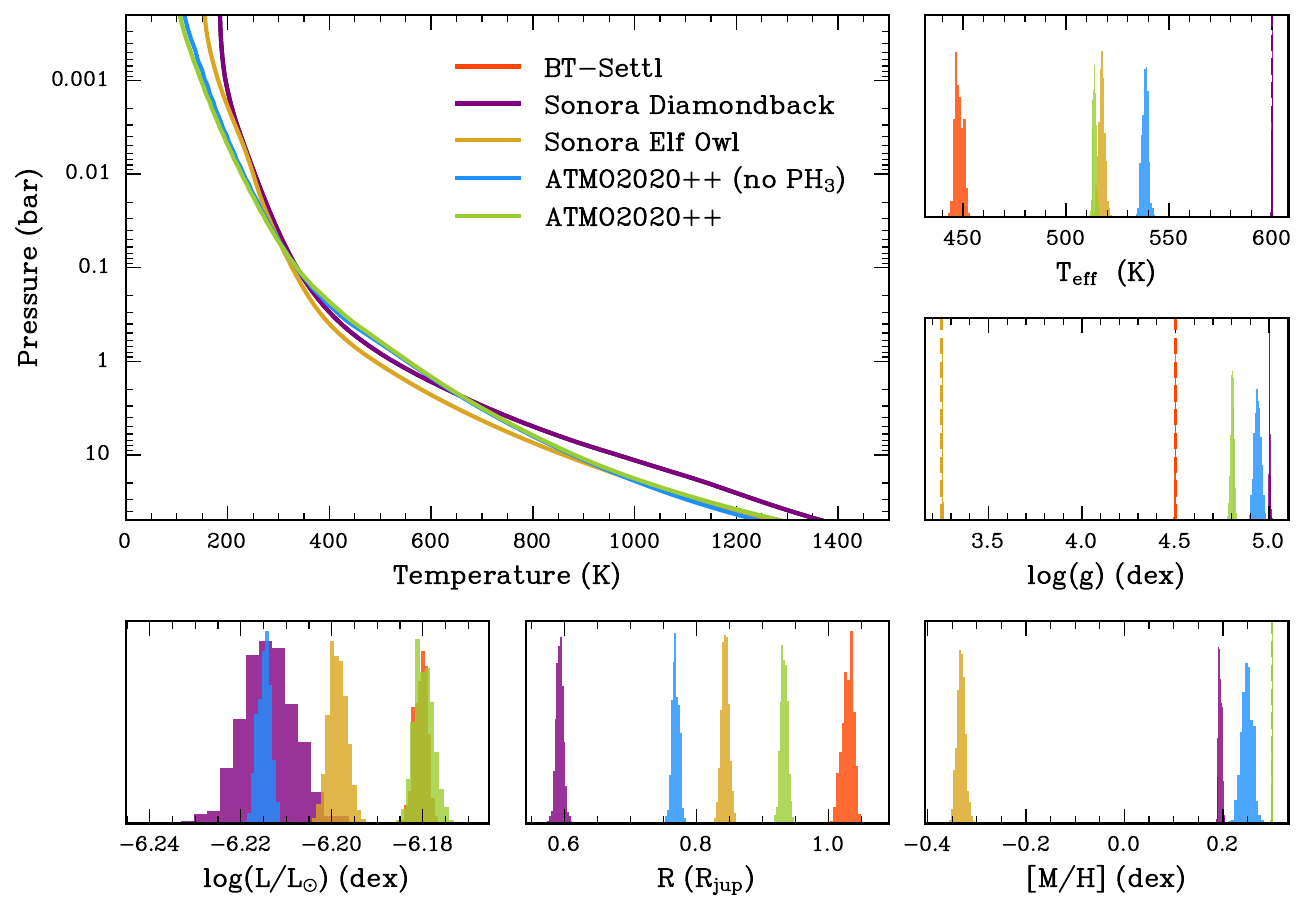}
    \caption{\small{\textit{Top-left panel:} Interpolated pressure–temperature (P–T) profiles for models providing a P–T grid (excluding BT-Settl) in Sect.~\ref{sec3.1}. \textit{Remaining panels:} Posteriors of key parameters. Vertical dashed indicate the boundaries of the respective model grids when encountered during the inversion.}}
    \label{fig1_comp_models_bis}
\end{figure*}

\setlength{\tabcolsep}{6pt}
\renewcommand{\arraystretch}{1.5}
\begin{table}[h!]
\tiny
    \centering
    \caption{\centering \small{Model comparison and significance criteria for all models of Sect.~\ref{sec3.1}.}}
	\begin{tabular}{lllll}
    \hline
    \hline
    Model & $\ln \mathcal{B}$ & $\Delta$AIC & $\Delta$BIC & $\Delta$BPICS \\
    \hline
    BT-Settl & 3381 & 6776 & 6751 & 6775 \\
    Sonora Diamondback & 2383 & 4767 & 4744 & 4768 \\
    Sonora Elf Owl & 331 & 665 & 668 & 656 \\
    ATMO2020++ (no PH$_3$) & 139 & 286 & 286 & 286 \\
    ATMO2020++ & 0 & 0 & 0 & 0 \\
    \hline
    \end{tabular}
    \tablefoot{\small{We use the formula from \cite{Spiegelhalter2002} to define the effective number of parameters in the computation of BPICS.}}
    \label{bayes_classic}
\end{table}

Fig. \ref{fig1_comp_models}, \ref{fig1_comp_models_bis}, Tables \ref{bayes_classic} and \ref{tab_parameters} summarize the inversion results using the five different models. While most models manage to reproduce the overall shape of the combined observations, we notice significant data–model mismatches across all instruments. We observe an important dispersion between models and their retrieved posterior distributions (see Fig. \ref{fig1_comp_models_bis}), with almost no overlap. This huge dispersion suggests the models are unable to robustly extract the physical parameters. Comparing the log-Bayes factors (see Table \ref{bayes_classic}), the ATMO2020++ models (with and without PH$_3$) are statistically preferred among those tested. This result is in line with previous atmospheric analysis of the target \citep{zhang_disequilibrium_2024}, although the model with PH$_3$ seems to be preferred ($\ln\mathcal{B}$~=~138). The PH$_3$ detection significance will be discussed in Sect. \ref{sec4.5}.

The retrieved effective temperatures range from $450$ to $540$~K for models that converged within their parameter grids (excluding Sonora Diamondback, which converges at the grid edge with \Teff~$>600$~K). This is broadly consistent with previous expectations from evolutionary tracks, which predict \Teff~=~$ 483^{+44}_{-53}$~K \citep{zhang_disequilibrium_2024}. The lower limit of \Teff~$>600$~K obtain with Sonora Diamondback probably explains the broader constrain observed in luminosity (see purple histogram in lower left panel of Fig.~\ref{fig1_comp_models_bis}) due to the degeneracy with the radius. We note that this temperature is highly inconsistent with evolutionary predictions. Surface gravity is poorly constrained, with retrieved values between $4.5$ and $5.0$~dex, significantly higher than the evolutionary estimate of log(g)~=~ $4.19^{+0.18}_{-0.13}$~dex. Notably, inversions using BT-Settl and Sonora Elf Owl reached the edges of their grids, with posteriors at $>4.5$~dex and $<3.25$~dex, respectively. However, comparisons between atmospheric and evolutionary log(g) estimates should be treated with caution, as they trace different physical processes (e.g. atmospheric structure versus radius contraction; \citealt{petrus_xshyne_2025}).

Except for Sonora Elf Owl, all grids that explored metallicity converged toward slightly super-solar values ([M/H]~$\in$~0.2–0.4~dex). Since we use the same dataset, it is reasonable to assume that the observed differences in metallicity between the various inversions arise from systematic offsets between the models. Because metallicity is correlated with log(g) \citep{zhang_uniform_2021}, the systematically higher surface gravities retrieved for most models in this setup may also bias the inferred metallicity. Moreover, metallicity is highly sensitive to the wavelength range and datasets included in the inversion \citep{petrus_jwst_2024}. Sonora Elf Owl is the only grid that also probes the carbon-to-oxygen ratio (C/O), yielding a sub-solar value of $0.317\pm0.005$.

Among the two grids that include cloud modeling, only Sonora Diamondback explores their effects explicitly through the sedimentation efficiency parameter of its silicate clouds (\fsed), which ranges from 1 (thick clouds) to 8 (thin clouds). Our best-fit value of \fsed~=~$3.6^{+0.1}_{-0.2}$ suggests a moderately cloudy atmosphere. However, since both cloudy models (BT-Settl and Sonora Diamondback) are among the least favored in our analysis, this constraint should be interpreted with caution, as it may not be strongly informative about the actual cloud content and properties of COCONUTS-2~b.

The vertical (eddy) diffusion coefficient \Kzz\;(in cm$^2$.s$^{-1}$) parametrizes the efficiency of atmospheric vertical mixing. In the Sonora Elf Owl grid, this parameter is explored logarithmically from $10^{2}$~cm$^2$.s$^{-1}$ (inefficient mixing) to $10^{9}$~cm$^2$.s$^{-1}$ (efficient mixing). We retrieve a value of \Kzz~=~$(9.8^{+0.5}_{-0.7})\times10^{3}$~cm$^2$.s$^{-1}$, indicative of moderate vertical mixing. The vertical mixing can also be retrieve analytically from the inferred log(g) using Fig. 1 of \cite{phillips_new_2020} for the two ATMO2020++ grids. Both ATMO2020++ models suggest a more vigorous vertical mixing with \Kzz~=~$(2.5\pm0.1)\times10^{5}$~cm$^2$.s$^{-1}$ and \Kzz~=~$(1.3\pm0.1)\times10^{5}$~cm$^2$.s$^{-1}$ for the model with and without PH$_3$ respectively. This strong vertical mixing can also be seeing on Fig.~\ref{mass_fraction_classic_vs_GP} where all key chemical abundances are quenched. However, as noted and explored in \citet{Mukherjee2022, mukherjee_sonora_2024}, \Kzz\;is expected to vary significantly with pressure, depending on whether the local atmospheric layer is radiative or convective. This is an effect that is not accounted for in this setup. In addition, the retrieved $\Kzz$ may partially encode viewing-geometry effects, further contributing to the degeneracy of this parameter \citep{tan_atmospheric_2021, suarez_ultracool_2023, petrus_xshyne_2025}. Recent studies have indeed investigated how mixing strength may vary in more complex 3D atmospheres \citep{Visscher2010, Mukherjee2022, Lacy2023}.

The top-left panel of Fig.~\ref{fig1_comp_models_bis} compares the retrieved pressure–temperature (P–T) profiles for Sonora Diamondback (purple), Sonora Elf Owl (yellow), and ATMO2020++ (green). All profiles agree well.

\subsection{GP analysis}\label{sec3.2}

\begin{figure}[t]
\centering
    \includegraphics[scale=0.35]{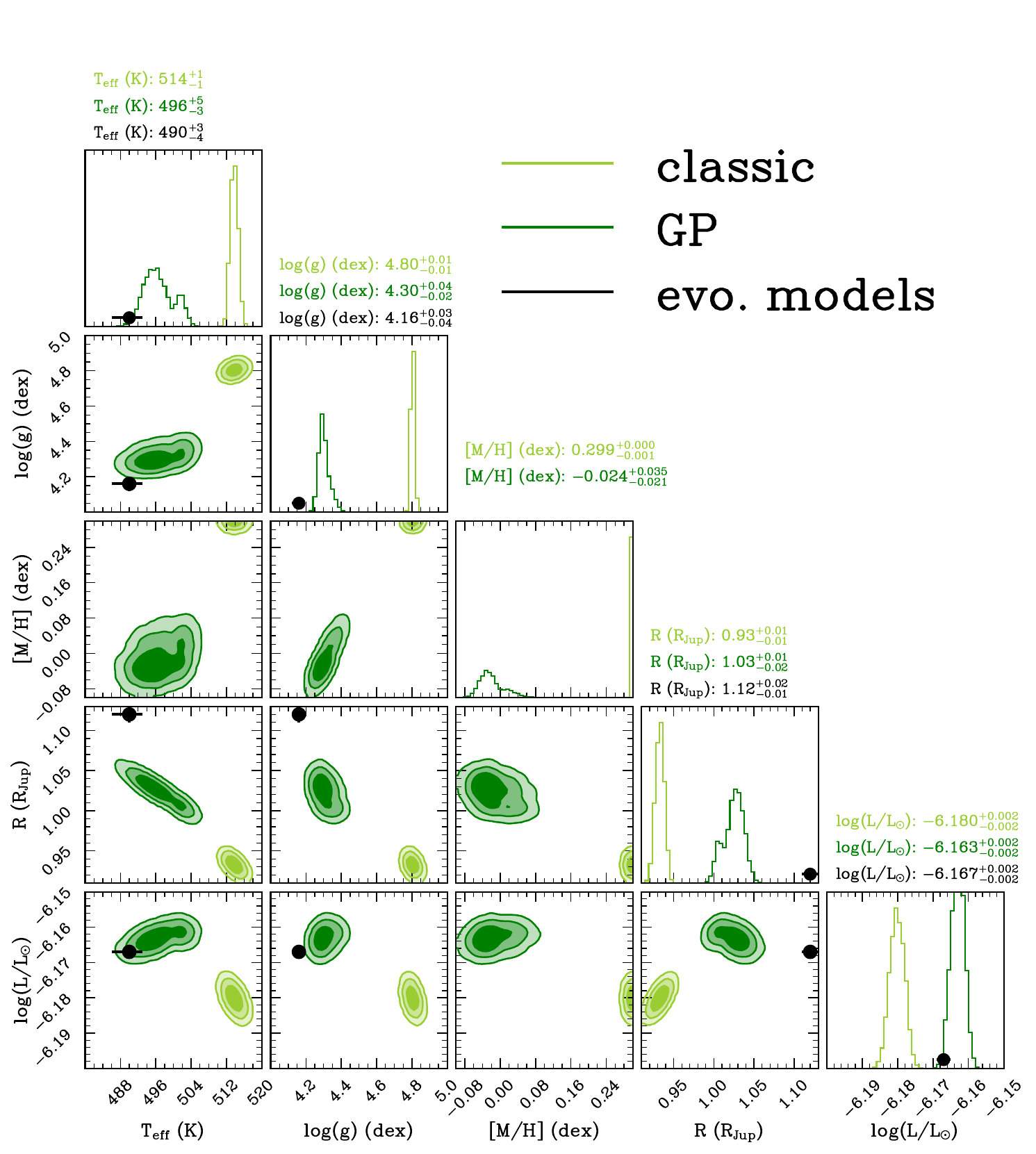}
    \caption{\small{Corner plot comparing forward modeling using the ATMO2020++ grid with (dark green) and without (light green) Gaussian Processes (GP). Black points indicate the mean predictions from evolutionary models with their associated error bars (see Sect.~\ref{sec4.6}).}}
    \label{fig2_corner_GPvsclassic}
\end{figure}

\setlength{\tabcolsep}{6pt}
\renewcommand{\arraystretch}{1.5}
\begin{table}[h!]
\tiny
    \centering
    \caption{\centering \small{Model comparison and significance criteria for all models of Sect.~\ref{sec3.2}.}}
	\begin{tabular}{lllll}
    \hline
    \hline
    Model & $\ln \mathcal{B}$ & $\Delta$AIC & $\Delta$BIC & $\Delta$BPICS\\
    \hline
    BT-Settl GP & 1219 & 2445 & 2426 & 2448 \\
    Sonora Diamondback GP & 732 & 1463 & 1457 & 1447 \\
    ATMO2020++ (no PH$_3$) GP & 319 & 643 & 630 & 648 \\
    ATMO2020++ GP & 94 & 197 & 123 & 196 \\
    Sonora Elf Owl GP & 0 & 0 & 0 & 0 \\
    \hline
    \end{tabular}
    \tablefoot{\small{We use the formula from \cite{Spiegelhalter2002} to define the effective number of parameters in the computation of BPICS.}}
    \label{bayes_GP}
\end{table}

In Sect.~\ref{sec3.1} we identified ATMO2020++ and its variation ATMO2020++ (no PH$_3$) as the two preferred models with ATMO2020++ (no PH$_3$) being the only model that does not converge outside its grid range for any parameter. Using the GP framework presented in Sect.~\ref{sec2.2}, we re-analysed the combined data set. Assuming each dataset is independent, we fit for each covariance matrix separately, resulting in three length-scales and three amplitudes as additional extra-grid parameters.

\begin{figure}[t]
    \centering
    \includegraphics[scale=0.58]{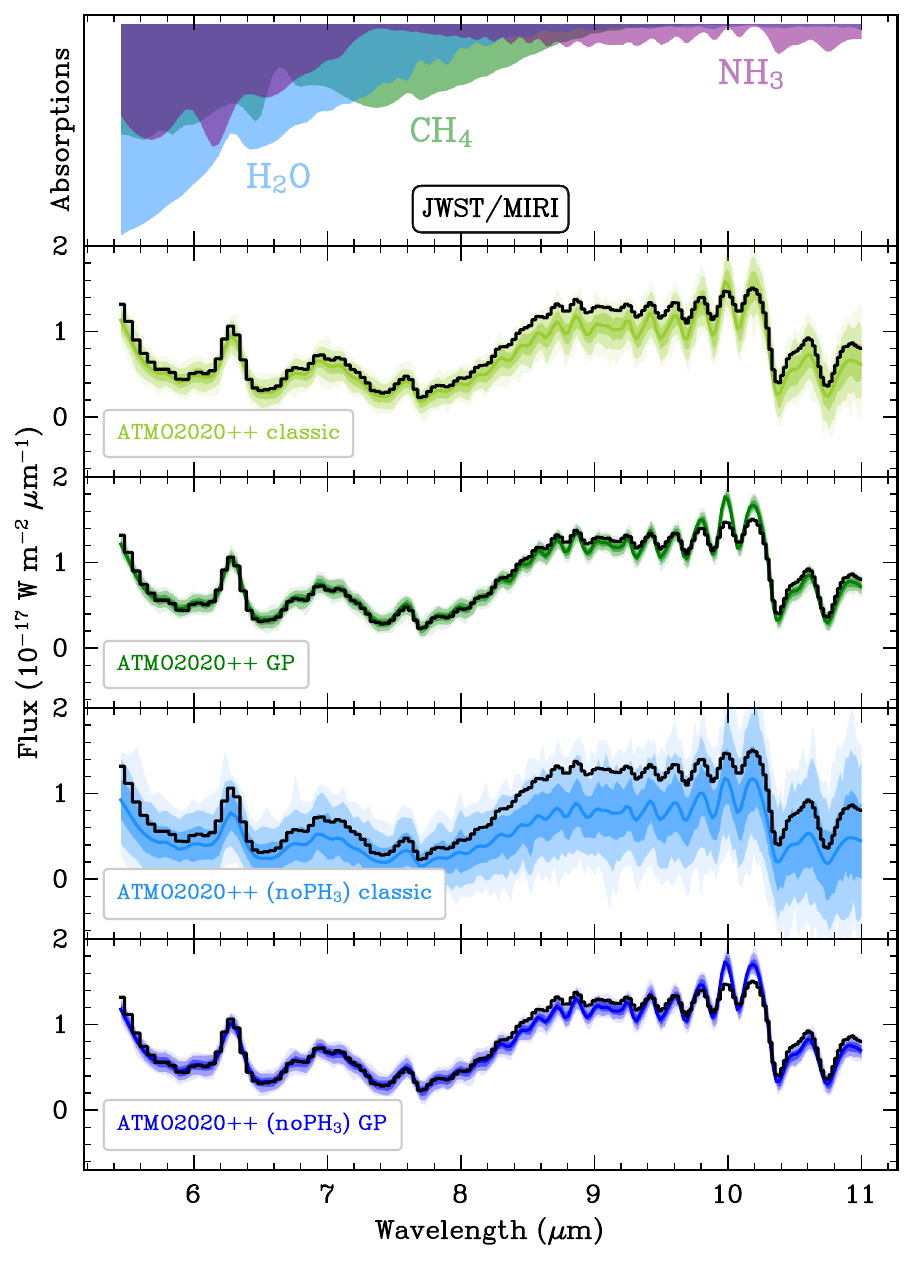}
    \caption{\small{Same as Fig.~\ref{fig_GPvsclassic_FLAMINGOS-2} but Comparison between the classical and GP-aided approaches on the JWST/MIRI-LRS spectrum using ATMO2020++ (in green) and ATMO2020++ (no PH$_3$, in blue). \textit{Top panel:} Main molecular absorption features calculated using \texttt{\textit{petitRADTRANS}}. \textit{Sub panels:} Black lines show the observed spectrum; colored lines show the corresponding model spectrum. Shaded regions around the model indicate the 1$\sigma$, 2$\sigma$, and 3$\sigma$ dispersions from 200 random draws from the retrieved covariance matrix $\vec{C}$. In the "classical" approach, the covariance matrix is assumed to be diagonal, $\vec{C} = \hat{s}^2\,\mathrm{diag}(\vec{\sigma}^2)$, where $\hat{s}^2$ is the noise scaling factor and $\vec{\sigma}$ the observational uncertainties. Zoom-in on the 5--11~$\mu$m region.}}
    \label{fig_GPvsclassic_MIRI}
\end{figure}

This section focuses on the two ATMO2020++ models; however, we still report the final parameters obtained with the other grids in Table~\ref{tab_parameters}, Table~\ref{tab_hyperparameters} and additional elements of discussion are provided in Appendix~\ref{app_othergrids}. Notably, based on the various significance criteria (Table~\ref{bayes_GP}), the Sonora Elf Owl models become preferred when GP are included, highlighting the importance of the GP framework in the modeling. However, these models still provide a comparatively poor fit, both visually (large residual offsets) and in terms of the inferred physical parameters (inconsistent with cooling model predictions). We therefore retain the focus of this section on the two ATMO2020++ grids. Final posteriors are compared with the classical approach and evolutionary models predictions in Fig.~\ref{fig2_corner_GPvsclassic} and summary plots can be found in Fig.~\ref{fig_GPvsclassic_FLAMINGOS-2}, \ref{fig_GPvsclassic_NIRSPec} and \ref{fig_GPvsclassic_MIRI}. In particular, in Fig.~\ref{fig_GPvsclassic_MIRI}, the GP approach significantly improves the fit to the MIRI-LRS observation, where the flux was previously underestimated. The effect is less noticeable for the other two spectra. In the first column of Table~\ref{tab_parameters}, both GP-aided results exhibit a significant decrease in the relative log-Bayes factor, suggesting that accounting for correlated noise is relevant for this target. In this configuration, the ATMO2020++ model including PH$_3$ is again favored over the version without it, with a relative log-Bayes factor of $\ln\mathcal{B}$~=~$319-94=225$ (see Table~\ref{bayes_GP}).

Focusing on ATMO2020++, Fig.~\ref{fig2_corner_GPvsclassic} shows both a significant shift and broadening of the posterior distributions between the classical and GP approaches for all atmospheric parameters. We obtain \Teff~$=496^{+5}_{-3}$~K and log(g)~$=4.30^{+0.04}_{-0.02}$~dex; consistent with predictions from evolutionary models at 0.9 and 3.0$\sigma$, respectively, using the values reported in Table~\ref{evolutionary_tracks} ($490^{+3}_{-4}$~K and $4.16^{+0.03}_{-0.04}$~dex; see Sect.~\ref{sec4.6} for a more detailed description). The metallicity shifts markedly, from a super-solar [M/H]~>~0.30~dex in the classical approach to solar [M/H]~$=-0.02^{+0.03}_{-0.02}$~dex with the GP model; probably linked to the associated decrease in log(g). From our inversion, we infer a radius of R~$=1.03^{+0.01}_{-0.02}$~\Rjup (consistent at 3.3$\sigma$), which translates into an estimated luminosity of log(L/L$_{\odot}$)=~$-6.163\pm0.002$~dex when combining the datasets. This luminosity estimate is also consistent with the value of ~$-6.162\pm0.006$~dex obtain by \citealp{zhang_disequilibrium_2024} with the same model. 

The retrieved hyperparameters for each observation listed in Table~\ref{tab_hyperparameters} are consistent between the two ATMO2020++ models. These hyperparameters are also consistent across the other model grids (see Table~\ref{tab_hyperparameters} and Appendix~\ref{app_othergrids}). Among the datasets, MIRI-LRS exhibits the largest fitted correlation amplitude, with log($a$)$_\mathrm{MIRI}$~=~$1.24\pm0.02$. 

The top row of Fig.~\ref{fig_covariances} presents a zoom-in on the three retrieved covariance matrices, normalized to highlight their noise structures. The correlated noise pattern is particularly visible in the FLAMINGOS-2 matrix with correlated noise extending across multiple spectral channels. Its block-like structure arises from the spectral windows used in the analysis. 

The middle row displays the corresponding mean radial profiles, overplotted with the AutoCorrelation Function (ACF) of the residuals. Comparing these profiles allows for an assessment of whether the retrieved correlation lengths adequately capture the actual noise structure in the data, assuming the model provides a good fit and the residuals are therefore dominated by noise. We find that the retrieved correlation lengths generally match this noise structure, although the correlation length may be slightly underestimated for the MIRI-LRS observation. Spectrographs with the finest wavelength sampling (FLAMINGOS-2 and NIRSpec) exhibit the largest correlation lengths, often spanning multiple pixels. In contrast, the retrieved correlation length for MIRI-LRS is nearly equal to the pixel size ($\sim0.02$~$\mu$m), suggesting more localized noise correlations.

The bottom row of Fig.~\ref{fig_covariances} compares the Power Spectral Density (PSD) of the residuals, the injected observational uncertainties, and GP realizations. At low spectral frequencies, the PSD of the residuals (in green) is well captured by the GP (in red) and white noise (in blue) is not sufficient to explain the residual noise. At higher frequencies, for both the FLAMINGOS-2 and NIRSpec spectrographs, the PSD of the residuals is below both the one of the injected uncertainties and GP noise suggesting that the adopted noise model is too conservative at very small spectral scales.

Using the GP-aided spectral analysis on the combined dataset, we adopt the following parameters for COCONUTS-2~b: \Teff~$=496^{+5}_{-3}$~K, log(g)~$=4.30^{+0.04}_{-0.02}$~dex, [M/H]~$=-0.02^{+0.03}_{-0.02}$~dex, R~$=1.03^{+0.01}_{-0.02}$~\Rjup and log(L/L$_{\odot}$)~$=-6.166\pm0.002$~dex (from Sect.~\ref{sec4.6}).

\subsection{SED analysis}\label{sec4.6}

In Sects.~\ref{sec3.1} and \ref{sec3.2}, we constrain the bolometric luminosity of COCONUTS-2~b by propagating the posterior distributions of its effective temperature and radius using the Stefan–Boltzmann law (see the last column of Table~\ref{tab_parameters}). This approach implicitly assumes that the bolometric flux can be accurately described by a single effective temperature. A more direct and less assumption-driven estimate can be obtained by directly integrating the full (SED).

Because the wavelength coverage is not continuous, this SED must be completed. Previous studies \citep[e.g.,][]{, filippazzo_fundamental_2015, petrus_xshyne_2025, kiman_diversity_2025} have extrapolated the missing regions using Wien’s approximation at short wavelengths and the Rayleigh–Jeans tail at long wavelengths. This approach avoids introducing explicit atmospheric model assumptions (systematics) and is well suited for relatively warm L/T objects with smooth spectral energy distributions.

However, for very cold objects such as COCONUTS-2~b, whose emergent spectra are strongly shaped by molecular absorption, the assumption of smooth blackbody-like behavior outside the observed range may be less accurate. In this work, we chose to use the posterior chain from our best-fit model in Sect.~\ref{sec3.2} (ATMO2020++ with GP) to generate a sample of model spectra over the widest possible wavelength range (0.2--30~$\mu$m). We then propagate this sample alongside the posterior distribution of the radius to derive a probability distribution for the bolometric luminosity. As previously, to account for potential model–data discrepancies, we also apply the corrective term given by equation (3) of \cite{zhang_disequilibrium_2024}, yielding:

\begin{equation}
\text{log(L/L}_{\odot}) =-6.166\pm0.002~\text{dex}.
\label{equlum}
\end{equation}

By repeating the procedure describe above on MIRI-LRS's wavelength range, we can estimate the contribution of the MIRI-LRS's observation to the total bolometric flux to be $41\%$. This value highlights the crucial role of the new dataset in constraining the bolometric luminosity of COCONUTS-2~b.

Finally, we use the ATMO2020++ \citep{phillips_new_2020}, Sonora Bobcat \citep{marley_sonora_2021} and Sonora Diamondback hybrids \citep{morley_sonora_2024} evolutionary models to rederive the physical properties of COCONUTS-2~b. We assume a Gaussian distribution for the luminosity based and the value we estimate (equ.~\ref{equlum}) and Gaussian distribution for the age, based on the predictions from \cite{kiman_diversity_2025} ($414\pm23$~Myr). The inferred parameters are summarized in Table~\ref{evolutionary_tracks}. While these parameters are fairly consistent across models, there is an observable difference in radius and effective temperature between the ATMO2020 models and Sonora Bobcat models (see Fig.~\ref{evolutionary_tracks_fig_comp}). After concatenating the chains for each parameter following the Monte Carlo sampling, evolution models predict that COCONUTS-2~b has \Teff~$=490^{+3}_{-4}$~K, log(g)~$=4.16^{+0.03}_{-0.04}$~dex, R~ $=1.12^{+0.02}_{-0.01}$~\Rjup and M~$=7.3\pm0.3$~\Mjup. These predictions are compared to the ones from atmospheric models in Fig.~\ref{fig2_corner_GPvsclassic}.

\section{Discussion}\label{sec4}

\subsection{Atmosphere and formation of COCONUTS-2~b}\label{sec4.1}

In Sect.~\ref{sec3.1}, we observed significant discrepancies in the atmospheric predictions from the different tested models. The cloud-free models (Sonora Elf Owl, ATMO2020++, and ATMO2020++ without PH$_3$) are significantly preferred based on Bayesian evidence, although difference between them are also noticeable in the overall fit (see Fig.~\ref{fig1_comp_models}).

In particular, in the Y-band (FLAMINGOS-2) and N-band (MIRI), all models underestimate the observed flux. As suggested in \cite{zhang_disequilibrium_2024}, the discrepancies observed in the Y-band may be due to uncertainties in the modeling of the pressure-dependent Na and K lines. Although the preferred and most recent grids we use (ATMO2020++ and Sonora Elf Owl) include a self-consistent decrease in the abundance of these chemical species in the upper atmospheric layers through condensation and precipitation processes, we argue that increased precipitation or additional processes are needed to reduce the abundance of Na/K to match the observations. The flux underestimation in MIRI’s wavelength range is not seen in the GP inversions (Sect.\ref{sec3.2}, Fig.\ref{fig_GPvsclassic_MIRI}), suggesting that ATMO2020++ can reproduce this part of the spectrum reasonably well without invoking additional physical or chemical processes. This discrepancy between the classical and GP approaches arises from differences in the treatment of the noise as, in the classical approach, the noise marginalization effectively reduces the importance of the MIRI-LRS data during the inversion (see Sect.~\ref{sec4.2}).

We pushed the atmospheric analysis using a GP-aided framework in Sect.~\ref{sec3.2} on the ATMO2020++ and ATMO2020++ (no PH$_3$) models which resulted in a significant improvement in both the fit quality and consistency with evolutionary predictions. For both best-fit models, we retrieved metallicities consistent with the metallicity of the primary ($0.00\pm0.08$~dex; \citealp{hojjatpanah_catalog_2019}) with [M/H]~$=-0.02^{+0.03}_{-0.02}$~dex and [M/H]~$=-0.09\pm0.02$~dex for the models including and excluding PH$_3$, respectively. This general agreement suggests that COCONUTS-2~b has not experienced significant enrichment through core erosion or planetesimal accretion, and is instead compatible with a binary-like formation scenario. However, [M/H] remain among the most degenerate parameters in our analysis analysis (Fig.~\ref{fig1_comp_models_bis} and Table~\ref{tab_parameters}), with retrieved values spanning from super- to sub-solar depending on the model employed. For instance, both Sonora models favor a more pronounced sub-solar metallicity with [M/H]~$=-0.43\pm0.01$~dex and [M/H]~$=-0.30\pm0.02$~dex for Diamondback and Elf Owl respectively when using GP (see Table~~\ref{tab_parameters}). This could in turn suggest a different formation pathway (planetary, late accretion, ...). Overall, this parameter being strongly coupled to the entire atmospheric composition means it is highly sensitive to the treatment of chemical processes, particularly non-equilibrium chemistry, which is nevertheless properly included in all favored models.

\subsection{Noise modeling}\label{sec4.2}

Accounting for all noise sources within a single inversion framework is a challenging task \citep[e.g.,][]{czekala_constructing_2015,gully-santiago_interpretable_2022}. One of the simplest and least computationally expensive approaches is to include a noise scaling parameter in the Bayesian estimator, which can either be fitted directly \citep[e.g.,][]{zhang_disequilibrium_2024} or marginalized over \citep[e.g.,][]{ruffio_radial_2019}. This scaling factor can be additive \citep[e.g.,][]{zhang_disequilibrium_2024} or, more commonly, multiplicative \citep[e.g.,][]{ruffio_radial_2019}, and aims to compensate for missing and uncorrelated noise sources.

Marginalizing over this factor has been shown to significantly improve the robustness of retrieved posteriors in both low- \citep{landman__2024, denis_characterization_2025} and high-S/N regimes \citep{de_regt_eso_2024, de_regt_eso_2025}. However, our study demonstrates that such an approach is less suited to deal with multi-modal datasets. Marginalizing over this scaling factor effectively “equalizes” the S/N contribution of each dataset. While MIRI-LRS intrinsically has the highest S/N and would therefore be expected to dominate the fit, the optimization process increases its noise-scaling factor (see Table~\ref{noise_scalings}), effectively reducing its weight and enhancing the contribution of the FLAMINGOS-2 and NIRSpec spectra. This effect is clearly visible in Fig.~\ref{fig_GPvsclassic_MIRI}: in the classical approach, the noise-scaling parameter allows the model to underestimate the MIRI-LRS flux, whereas this bias is not present in the GP approach, which does not marginalize over the noise-scaling factor.

In this study, we used GPs to account for unmodeled spectral correlations, fitting for correlation lengths and amplitudes for each spectroscopic observations. We assessed the reliability of the retrieved correlation lengths from the ACF and PSD of the residuals (see Sect.~\ref{sec3.2} and Fig.~\ref{fig_covariances}). In contrast, the correlation amplitudes depend on the overall scale of the residuals; their reliability is therefore best evaluated through injection tests as illustrated in Appendix~\ref{app_inv}. 

Accounting for correlated noise in both ATMO2020++ models both statistically improved the fit quality and the consistency of most of the retrieved parameters (Table~\ref{tab_parameters}). These correlations may arise from a combination of instrumental effects, and model systematics. In practice, the two contributions are difficult to disentangle from the residuals alone, since the empirical ACF and PSD reflects both (residuals = observation - model). For the FLAMINGOS-2 and NIRSpec instruments, both the residual autocorrelation and the GP reveal correlation lengths larger than (or in the order of) the instrumental Line Spread Function ($\sim$2–3 pixels, see middle row of Fig.~\ref{fig_covariances}), indicating the presence of additional, more complex noise structures in the data that must be accounted for in the analysis.

More broadly, GP applications in exoplanet atmospheric characterization can extend beyond noise modeling. They have, for instance, been extensively used in transit spectroscopy to model systematic noise sources and correct for stellar contamination \citep[e.g.,][]{gibson_gp_2012}. In the context of forward modeling, they can also be employed to remove local deviant spectra using the overal trends of the grids, therefore mitigating the effect of these spectra on the posterior distributions \citep{czekala_constructing_2015}. Moreover, GPs can propagate interpolation uncertainties into the final posteriors, as demonstrated in \cite{czekala_constructing_2015} with their \texttt{\textit{Starfish}}\footnote{\url{https://starfish.readthedocs.io/en/latest/}} framework. These last two aspects, while promising, were not explored in the present work. Future forward-modeling efforts will benefit from incorporating these GP-based developments, both to better characterize the noise structure and to more robustly disentangle observational systematics from model-driven discrepancies.

\subsection{PH$_3$ analysis}\label{sec4.5}

Phosphine (PH$_3$) was expected to be the primary phosphorous molecule in the low-temperature atmospheres of brown dwarfs and giant exoplanets \citep{fegley_atm_1996, visscher_atm_2006}. However, most observational searches for PH$_3$ in these planetary-mass objects \citep[e.g.,][]{morley_an_2018, wallack_compas_2024} have provided only upper limits $\sim$100 times lower than the abundances predicted by atmosphere models \citep{beiler_tale_2024}. In their paper \citealp{beiler_tale_2024}, discussed atmospheric mechanisms that could explain the observed underabundance of PH$_3$, including a vertical eddy diffusion coefficient that varies with altitude, incorrect chemical pathways, elements condensing out in forms such as NH$_4$H$_2$PO$_4$, or wrong quenching approximations.

Throughout our analysis, the ATMO2020++ including PH$_3$ remained statistically preferred compared to the same model without this molecule. PH$_3$ strongest features are located between 4 and 4.5~$\mu$m with a deep, broad line at 4.3~$\mu$m. This deep 4.3~$\mu$m feature is clearly visible on the final fit (second and third panels from the top of Fig.~\ref{fig_GPvsclassic_NIRSPec}) but does not appear in the data at this location, clearly suggesting that PH$_3$ is indeed absent from the observation. On the other hand, the model without PH$_3$ manages to fit this region (fourth and fifth panels from the top of Fig.~\ref{fig_GPvsclassic_NIRSPec}) but consistently overestimate the flux between 4~$\mu$m and 4.2~$\mu$m (with or without GP) where CH$_4$ lines are present.

To evaluate whether this broader 4--4.2~$\mu$m region could be driving the preference for the model with PH$_3$, we masked it and re-ran the inversions. In this case, ATMO2020++ without PH$_3$ was favored with $\ln\mathcal{B}$~=~724 using the classical approach and $\ln\mathcal{B}$~=~54 with the GP-aided approach. These results clearly indicate that the putative detection of PH$_3$ absorptions is highly sensitive to wavelength coverage and model systematics. Neither model successfully reproduces the NIRSpec data in this region implying that robust constraints on PH$_3$ detection are difficult to obtain with this approach.

To complement this analysis, we perform a cross-correlation exploration of the NIRSpec spectrum using molecular templates. Templates for H$_2$O, CO, CH$_4$, CO$_2$, NH$_3$, and PH$_3$ are generated with \texttt{\textit{petitRADTRANS}}\footnote{\url{https://petitradtrans.readthedocs.io/en/latest/}} \citep{Molliere2019, Blain2024} using the median temperature and chemical profiles from the ATMO2020++ (GP) inversion (see Fig.~\ref{mass_fraction_classic_vs_GP}). We used the line-by-line opacity computation mode to generate spectra with only one molecule at a time. Both the observed spectrum and the templates are continuum-removed and normalized, and the template spectral resolution is matched to that of the observations. To avoid the 3.6--3.8~$\mu$m gap, we perform the cross-correlation on the two NIRSpec detectors separately. The resulting cross-correlation functions (CCF) are shown in Fig.~\ref{ccf_all}. H$_2$O, CO, CH$_4$, CO$_2$ and NH$_3$ are detected at significance levels above 4$\sigma$, while no significant signal is recovered for PH$_3$, further supporting its absence (or reduced abundance) in the atmosphere of COCONUTS-2~b.

\section{Conclusion}\label{sec5}

In this work, we investigated the T9 planetary-mass companion COCONUTS-2~b.
Using new MIRI-LRS observations in combination with existing spectro-photometric data within a self-consistent, GP-aided framework, we refined the atmospheric properties of this object. From the GP-aided model analysis, we adopt the following parameters: \Teff~$=496^{+5}_{-3}$~K, log(g)~$=4.30^{+0.04}_{-0.02}$~dex, [M/H]~$=-0.02^{+0.03}_{-0.02}$~dex, and R~$=1.03^{+0.01}_{-0.02}$~\Rjup. These values are consistent with previous results (\citealp{zhang_second_2021,zhang_disequilibrium_2024,kiman_diversity_2025}) and with updated predictions from evolutionary models. The adopted metallicity, [M/H]~$=-0.02^{+0.03}_{-0.02}$~dex, is consistent with that of the primary, supporting a binary-like formation scenario for COCONUTS-2~b. The wavelength extension provided by the MIRI-LRS observations contribute 41\% of the bolometric flux. It enables a robust new constraint on the luminosity of COCONUTS-2~b: log(L/L$_{\odot}$)~$=-6.166\pm0.002$~dex. 

However, systematic discrepancies between the data and atmospheric models in specific bands (Y and N), as well as inconsistencies in the retrieved metallicities and surface gravities across different model grids, remain the main limitations to provide a fully robust characterization of the object. These issues point to missing or incomplete chemistry in current forward models and highlight the need to incorporate improved alkali condensation and rainout, potentially more complex cloud structures, and more generally, diabatic processes. Such processes are increasingly being recognized as essential for interpreting the cold brown dwarf population revealed by JWST (e.g., \citealp{faherty_methane_2024, merchan_diversity_2025, kiman_diversity_2025}; \textcolor{blue}{Biller et al. in prep.}). Accounting for them self-consistently will thus be equally important in preparation for the population of temperate, near–water–ice–line planets expected from GAIA DR4 and the ELT; and, in the longer term, for extending these modeling efforts toward exo-earths analogues.

We also find that “classical” noise-scaling schemes are ill-suited for heterogeneous, multi-modal datasets with varying resolution and S/N. In contrast, parametric covariance models provide a robust alternative for low- to moderate-size observation samples.

Future retrieval analyses will aim to constrain individual molecular abundances and explore more complex cloud prescriptions (\textcolor{blue}{Copeland et al., in prep.; Kühnle et al., subm.}) using the data presented here and with the upcoming MIRI–MRS observations, providing further insight into the formation history of this object.
Additionally, the high S/N MIRI-LRS spectrum will enable constraints on the isotope ratios (e.g., D/H, $^{12}$C/$^{13}$C).

\begin{acknowledgements}
      This work is heavily supported by a large collaboration of international people: Atmospheric modeling groups represented by P. Tremblin and M. W. Phillips (ATMO), F. Allard (deceased) (BT-Settl), J. J. Mang and C. V. Morley (Sonora). Collaborators who provided us with the data are represented by the M. Bonnefoy (MIRI-LRS), J. K. Faherty (NIRSpec), Z. Zhang (FLAMINGOS-2). This project has received funding from \emph{Agence Nationale de la Recherche} (ANR) under grant ANR-23-CE31-0006-01 (MIRAGES). This project was provided with computing HPC and storage resources by GENCI at TGCC thanks to the grant 2024-15722 and 2025-15722 on the supercomputer Joliot Curie’s SKL and ROME partition. This work is based [in part] on observations made with the NASA/ESA/CSA James Webb Space Telescope. The data were obtained from the Mikulski Archive for Space Telescopes at the Space Telescope Science Institute, which is operated by the Association of Universities for Research in Astronomy, Inc., under NASA contract NAS 5-03127 for JWST. These observations are associated with GTO program 2124 and 3514. This research has made use of the NASA Exoplanet Archive, which is operated by the California Institute of Technology, under contract with the National Aeronautics and Space Administration under the Exoplanet Exploration Program. G.-D.M.\ acknowledges the partial support of the Deutsche Forschungsgemeinschaft (DFG) through grant ``MA~9185/2-1''. This publication is based upon work from COST Action CA22133 ``PLANETS'' (\url{https://costactionplanets.github.io}), supported by COST (European Cooperation in Science and Technology).
      
\end{acknowledgements}

\bibliographystyle{aa}
\bibliography{COCONUTS2b_paper}

\begin{appendix}

\section{Additional figures and tables}

\begin{figure}[t]
    \centering
    \includegraphics[scale=0.58]{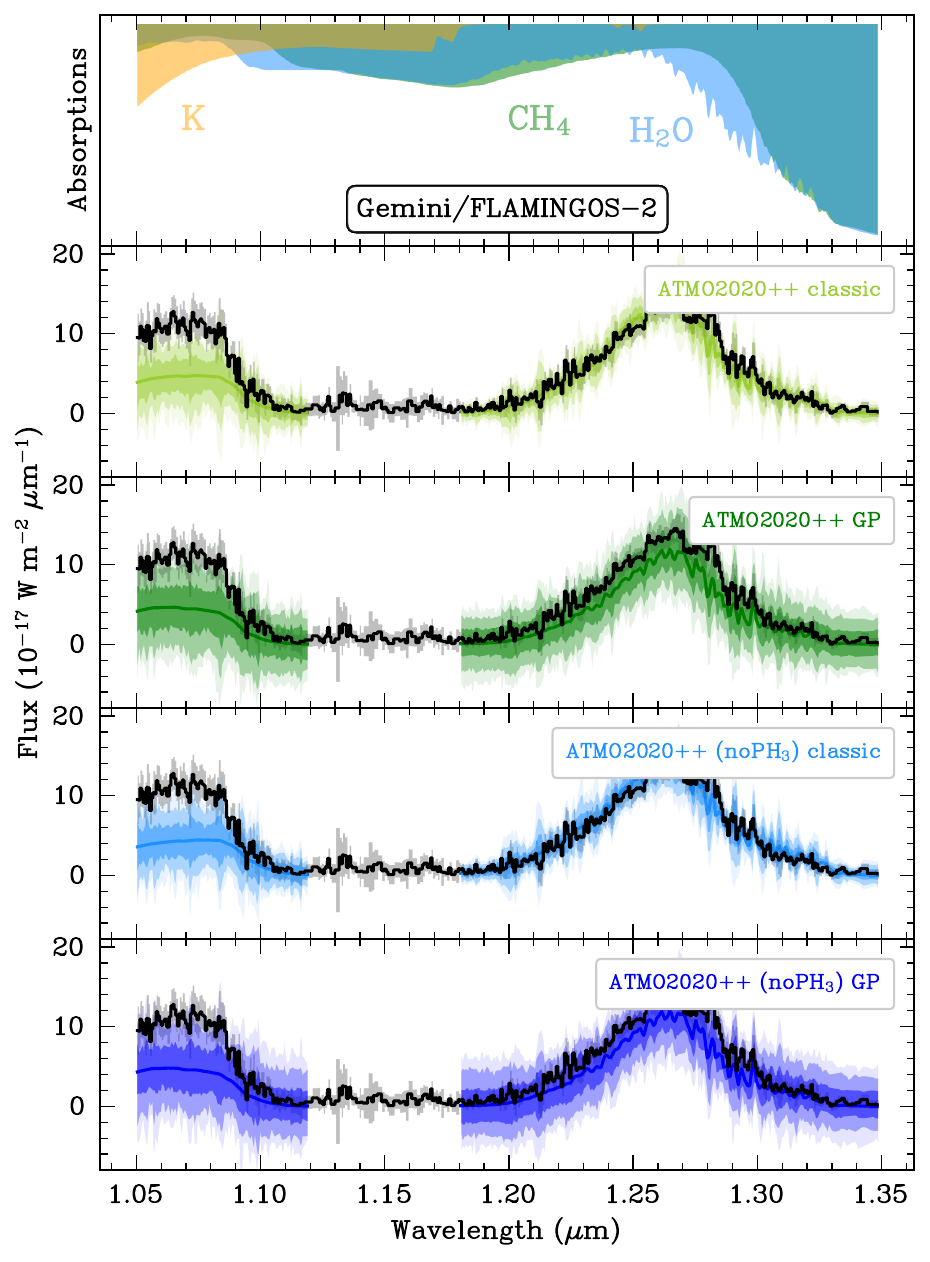}
    \caption{\small{Same as Fig.~\ref{fig_GPvsclassic_MIRI} but with Gemini/FLAMINGOS-2. Zoom-in on the Y and J bands.}}
    \label{fig_GPvsclassic_FLAMINGOS-2}
\end{figure}

\begin{figure}[t]
    \centering
    \includegraphics[scale=0.58]{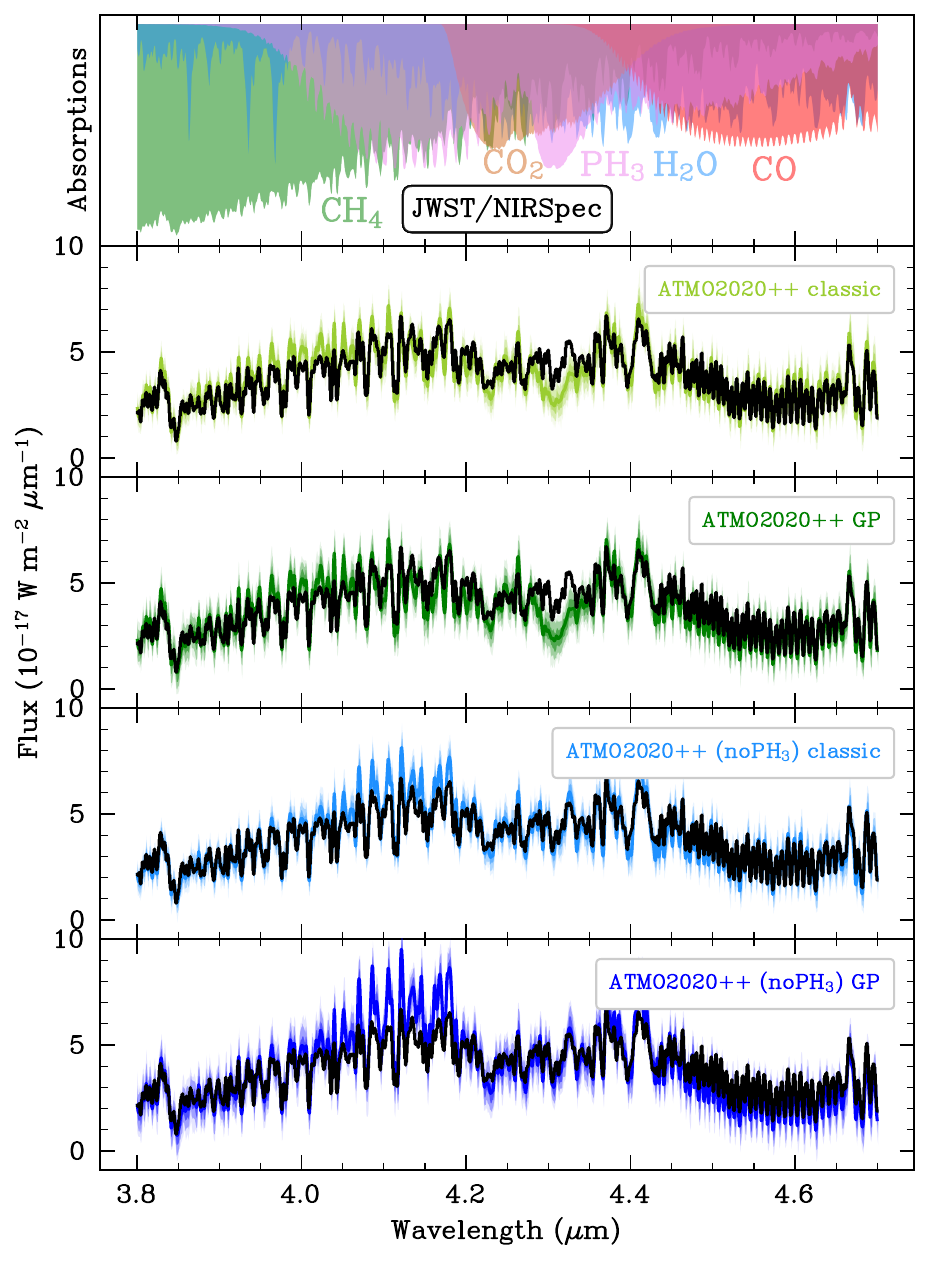}
    \caption{\small{Same as Fig.~\ref{fig_GPvsclassic_MIRI} but with JWST/NIRSpec. Zoom-in on the 3.8--4.7~$\mu$m region.}}
    \label{fig_GPvsclassic_NIRSPec}
\end{figure}

\begin{figure*}[t]
    \centering
    \includegraphics[scale=0.68]{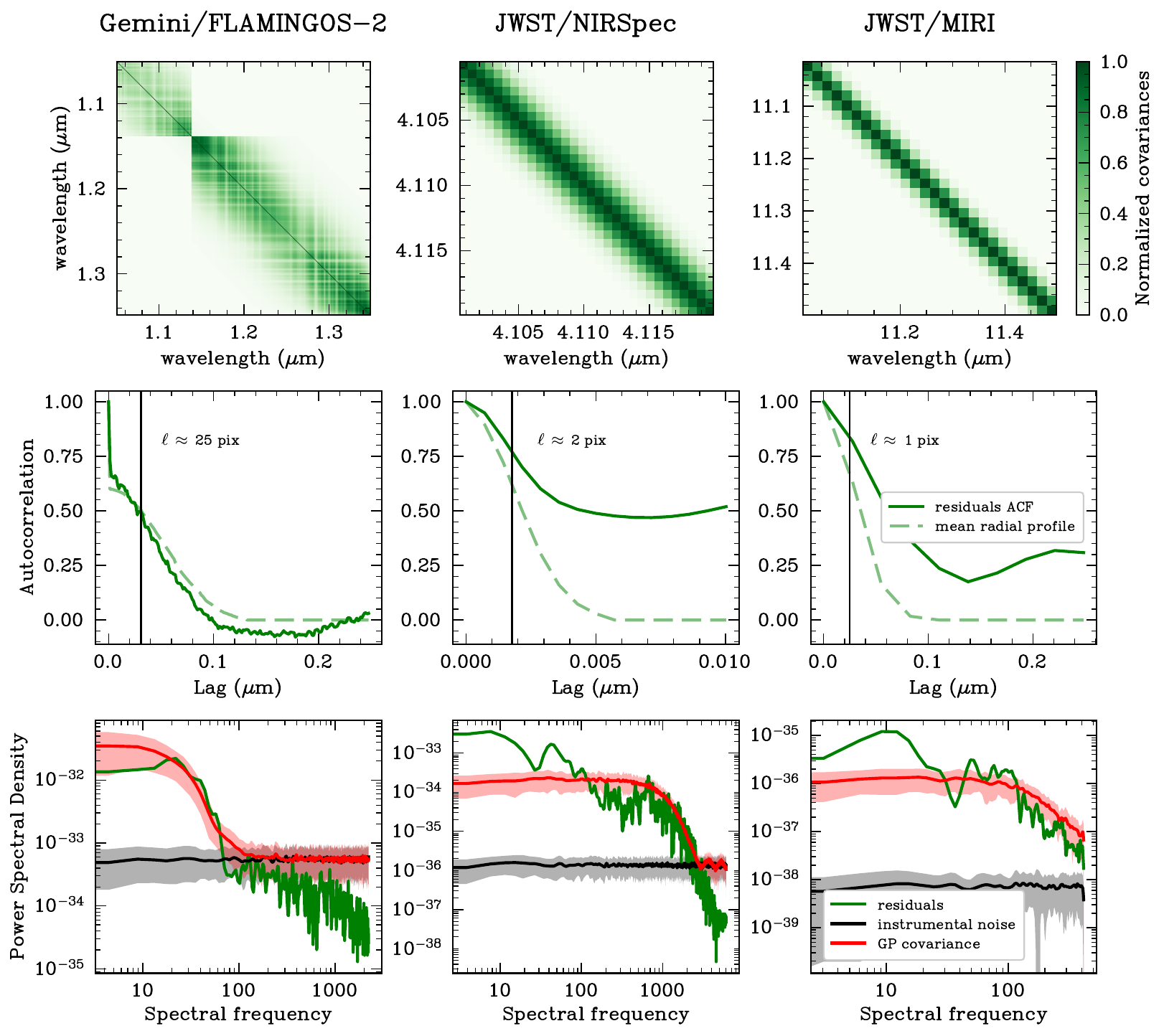}
    \caption{\small{GP diagnostics. \textit{Top row:} Zoom-in on the retrieved (normalized) covariance matrices from the GP-aided inversion using the ATMO2020++ model. The block-diagonal structure reflects the discontinuous spectral windows of the Gemini/FLAMINGOS-2 observations (see Section~\ref{sec3}). \textit{Middle row:} Autocorrelation of the residuals (data – model) in green, with the corresponding GP mean radial profile overplotted as a dotted green line. Vertical black lines indicate the GP correlation length-scales. \textit{Bottom row:} Power spectral densities (PSD) of the residuals, compared with random realizations from the injected instrumental noise (blue, $\pm1\sigma$ shaded) and from the fitted GP covariance (red, $\pm1\sigma$ shaded).}}
    \label{fig_covariances}
\end{figure*}

\begin{figure*}[t]
    \centering
    \includegraphics[scale=0.9]{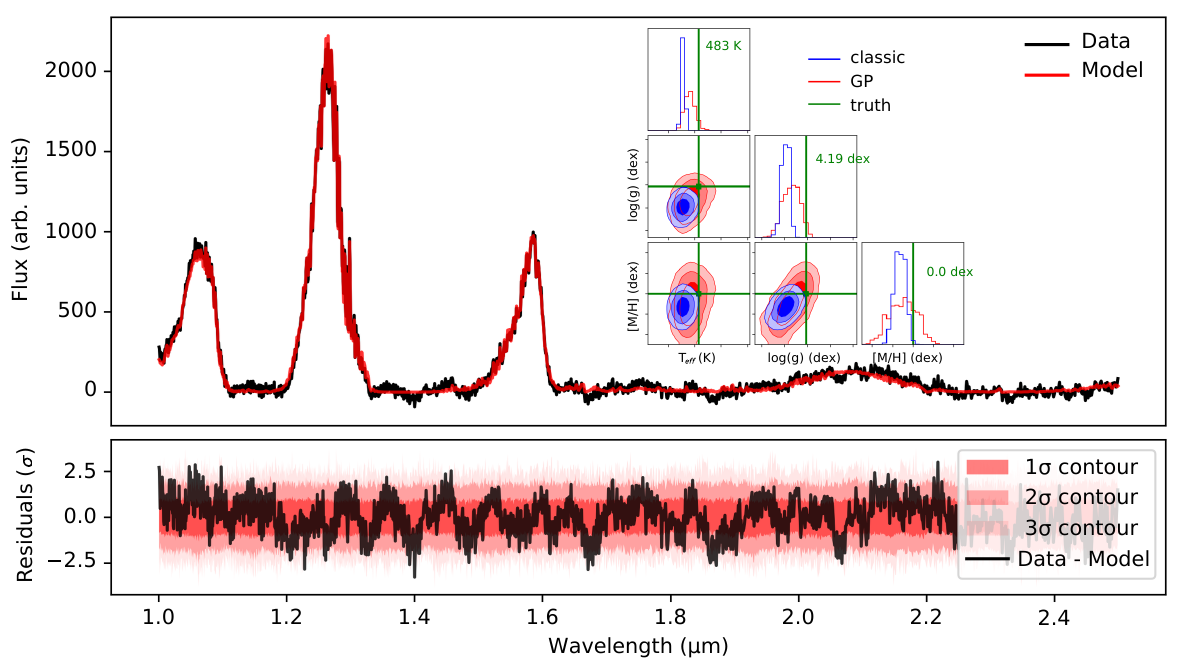}
    \caption{\small{Injection test summary figure. \textit{Top panel:} Comparison between the mock spectrum (black) and the fitted GP-aided model (red). \textit{Top sub-panel:} Corner plot of the grid parameters from the inversions with (red) and without (blue) GP. Green lines indicate the injected values. \textit{Bottom panel:} Normalized residuals between the mock data and the fitted GP-aided model (black). Shaded regions represent the 1$\sigma$, 2$\sigma$, and 3$\sigma$ dispersions from 200 random draws from the covariance matrix $C$.}}
    \label{fig_inj_test}
\end{figure*}

\begin{figure}[h!]
\centering
    \includegraphics[scale=0.6]{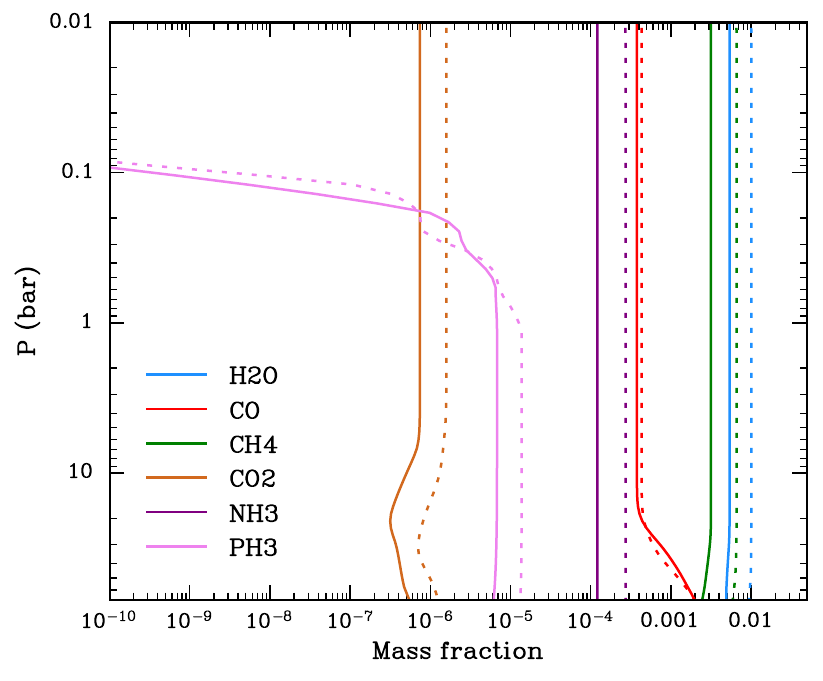}
    \caption{\small{Key species mass fractions interpolated from ATMO2020++. Dashed lines correspond to the classical inversion, while solid lines correspond to the GP-aided inversion.}}
    \label{mass_fraction_classic_vs_GP}
\end{figure}

\begin{figure}[h!]
\centering
    \includegraphics[scale=0.5]{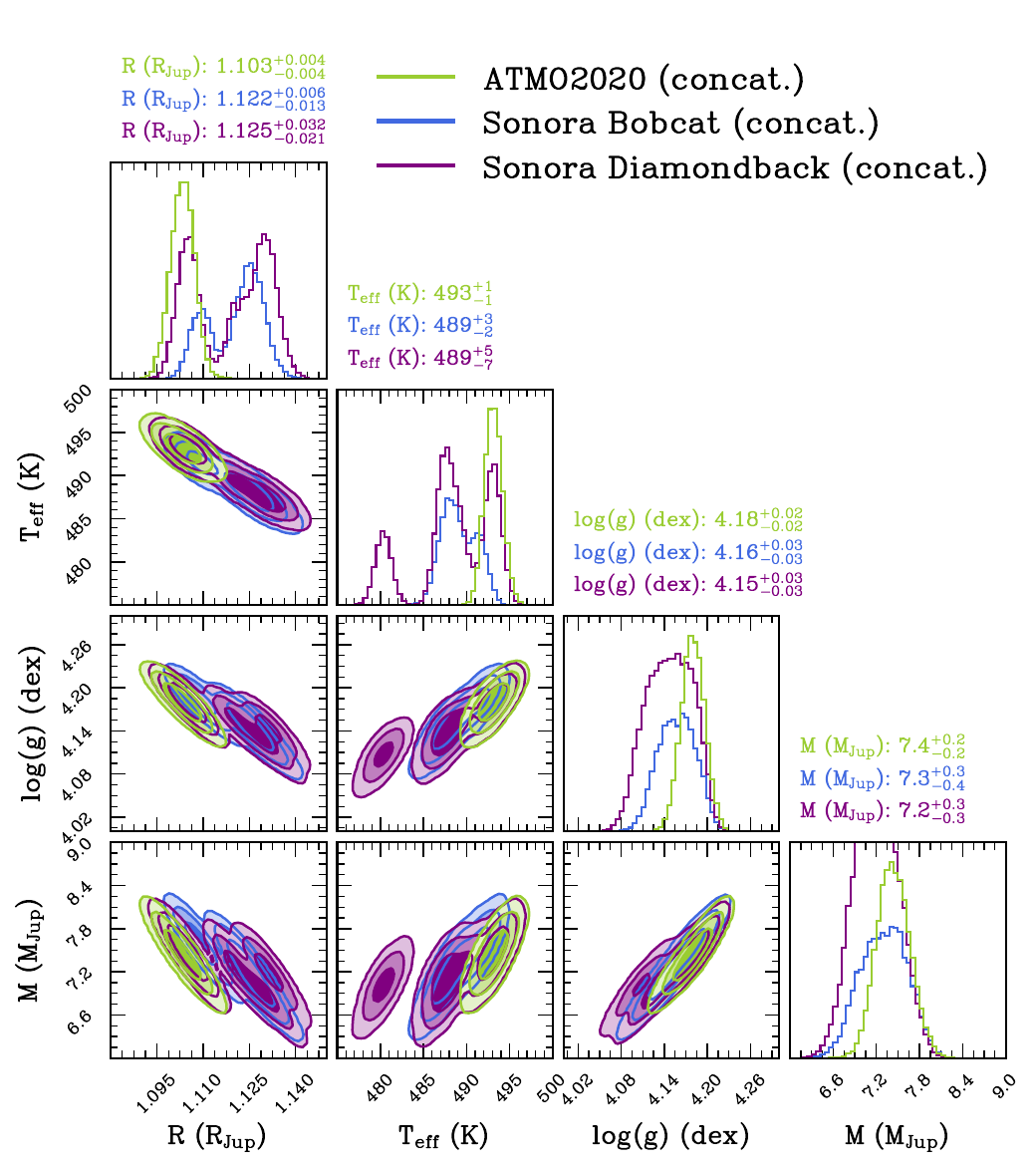}
    \caption{\small{Corner plot comparing the three families of evolutionary models (ATMO2020 in green, Sonora Bobcat in blue and Sonora Diamondback in purple). The secondary bumps in the Sonora distributions usually corresponds to the different [M/H] nodes. }}
    \label{evolutionary_tracks_fig_comp}
\end{figure}

\begin{figure}[h!]
\centering
    \includegraphics[scale=0.65]{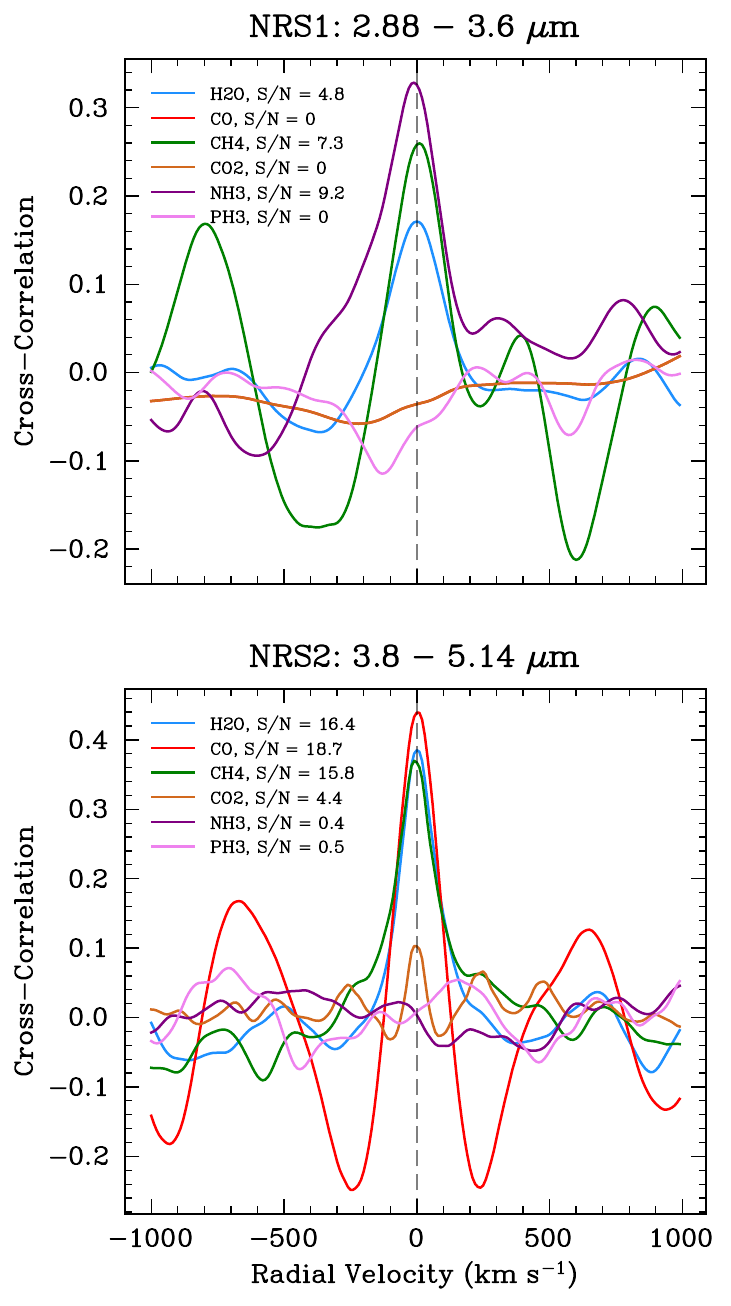}
    \caption{\small{Cross-correlation functions of COCONUTS-2 b NIRSpec spectra (at full resolution) with H$_2$O, CO, CH$_4$, CO$_2$, NH$_3$ and PH$_3$  molecular templates generated using \texttt{\textit{petitRADTRANS}}. The S/N is computed using formula (1) of \cite{houlle_direct_2021}. \textit{Upper pannel:} CCFs obtained using the NRS1 detector where H$_2$O, CH$_4$ and NH$_3$ are detected. \textit{Lower pannel:} CCFs obtained using the NRS2 detector where H$_2$O, CO, CH$_4$ and CO$_2$ are detected.}}
    \label{ccf_all}
\end{figure}

\setlength{\tabcolsep}{6pt}
\renewcommand{\arraystretch}{1.5}
\begin{table}[h!]
\tiny
    \centering
    \caption{\centering \small{Noise scaling factors ($\hat{s_i} = \frac{\chi_i^2}{N_i}$) from the "classical" inversions of Sect.~\ref{sec3.1}.}}
	\begin{tabular}{llll}
    \hline
    \hline
    Model & $\hat{s}_\text{FLAMINGOS-2}$ & $\hat{s}_\text{NIRSpec}$ & $\hat{s}_\text{MIRI}$ \\
    \hline
    BT-Settl & 1.63 & 118.44 & 48715.21 \\
    Sonora Diamondback & 1.24 & 70.91 & 33987.07 \\
    Sonora Elf Owl & 1.00 & 21.97 & 18858.41 \\
    ATMO2020++ (no PH$_3$) & 1.03 & 20.84 & 2422.83 \\
    ATMO2020++ & 1.04 & 19.83 & 11812.63 \\
    \hline
    \end{tabular}
    
    \label{noise_scalings}
\end{table}

\setlength{\tabcolsep}{15pt}
\renewcommand{\arraystretch}{1.5}
\begin{table*}[h!]
\tiny
    \centering
    \caption{\centering \small{COCONUTS-2b properties derived from the different evolutionary models explored.}}
	\begin{tabular}{lcccc}
    \hline
    \hline
    Model & \Teff (K) & log(g) (dex) & R  (\Rjup) & M (\Mjup) \\
    \hline
    ATMO2020 CEQ & $493\pm1$ & $4.18\pm0.02$ & $1.103\pm0.004$ & $7.4\pm0.2$ \\
    ATMO2020 NEQ weak & $493\pm1$ & $4.18\pm0.02$ & $1.103\pm0.004$ & $7.4\pm0.2$ \\
    ATMO2020 NEQ strong & $493\pm1$ & $4.18\pm0.02$ & $1.103\pm0.004$ & $7.4\pm0.2$ \\
    Sonora Bobcat [M/H]~=~-0.5~dex & $491\pm1$ & $4.18\pm0.02$ & $1.109\pm0.004$ & $7.6\pm0.2$ \\
    Sonora Bobcat [M/H]~=~0.0~dex & $488\pm1$ & $4.16^{+0.01}_{-0.02}$ & $1.123\pm0.004$ & $7.3\pm0.2$ \\
    Sonora Bobcat [M/H]~=~+0.5~dex & $487\pm1$ & $4.13\pm0.02$ & $1.128\pm0.004$ & $7.0\pm0.2$ \\
   Sonora Diamondback "hybrid"  [M/H]~=~-0.5~dex & $493\pm1$ & $4.18\pm0.02$ & $1.105\pm0.004$ & $7.5\pm0.2$ \\
   Sonora Diamondback "hybrid"  [M/H]~=~0.0~dex & $489\pm1$ & $4.15\pm0.02$ & $1.121^{+0.004}_{-0.003}$ & $7.2\pm0.2$ \\
   Sonora Diamondback "hybrid"  [M/H]~=~+0.5~dex & $488\pm1$ & $4.13\pm0.02$ & $1.130\pm0.004$ & $6.9\pm0.2$ \\
   Sonora Diamondback "hybrid-grav"  [M/H]~=~-0.5~dex & $493\pm1$ & $4.18\pm0.02$ & $1.105\pm0.004$ & $7.4\pm0.2$ \\
   Sonora Diamondback "hybrid-grav"  [M/H]~=~0.0~dex & $487\pm1$ & $4.14\pm0.02$ & $1.131^{+0.004}_{-0.003}$ & $7.2\pm0.2$ \\
   Sonora Diamondback "hybrid-grav"  [M/H]~=~+0.5~dex & $480\pm1$ & $4.11\pm0.02$ & $1.164\pm0.004$ & $7.0\pm0.2$ \\
    \hline
    \hline
    Concatenated & $490^{+3}_{-4}$ & $4.16^{+0.03}_{-0.04}$ & $1.12^{+0.02}_{-0.01}$ & $7.3\pm0.3$ \\ 
    \hline
    \end{tabular}
    \tablefoot{\small{“Concatenated” refers to the values and constraints obtained by merging the parameter chains after the Monte-Carlo exploration.}}
    \label{evolutionary_tracks}
\end{table*}

\setlength{\tabcolsep}{2pt}
\renewcommand{\arraystretch}{1.5}
\begin{table*}[h!]
\tiny
    \centering
    \caption{\centering \small{Inversion results of Sect. \ref{sec3}.}}
	\begin{tabular}{lccccccccc}
    \hline
    \hline
    Parameter & $\ln \mathcal{B}$ & \Teff & log(g) & [M/H] & C/O & f$_\text{sed}$ & log(K$_\text{zz}$) & R & log(L/L$_{\odot}$)\\
    Units & & (K) & (dex) & (dex) & & & log(cm$^2$.s$^{-1}$) & (\Rjup) & (dex) \\
    \hline
    BT-Settl priors & & $U(200, 1000)$ & $U(3.5, 4.5)$ & & & & & $U(0, 2)$ & \\
    BT-Settl posteriors & & & & & & & & & \\
    classic & 7643 & $447^{+3}_{-2}$ & $>4.49$ & (0.00) & (0.55) & (microphys.) & (profile) & $1.03\pm0.01$ & $-6.180^{+0.001}_{-0.002}$ \\
    GP & 1219 & $463\pm3$ & $4.48^{+0.01}_{-0.02}$ & (0.00) & (0.55) & (microphys.) & (profile) & $0.96\pm0.01$ & $-6.190\pm0.003$ \\
    \hline
    Sonora Diamondback priors & & $U(400, 600)$ & $U(3.5, 5.5)$ & $U(-0.5, 0.5)$ & & $U(1, 8)$ & & $U(0, 2)$ & \\
    Sonora Diamondback posteriors & & & & & & & & & \\
    classic & 6645 & $>600$ & $5.000^{+0.003}_{-0.002}$ & $0.194\pm0.004$ & (0.458) & $3.6^{+0.1}_{-0.2}$ & (profile) & $0.59\pm0.01$ & $-6.214\pm0.006$ \\
    GP & 732 & $515\pm1$ & $4.500^{+0.003}_{-0.004}$ & $-0.43\pm0.01$ & (0.458) & $>7.1$ & (profile) & $1.65^{+0.02}_{-0.03}$ & $-5.71\pm0.01$ \\
    \hline
    Sonora Elf Owl priors & & $U(400, 600)$ & $U(3.25, 5.5)$ & $U(-1.0, 1.0)$ & $U(0.229, 1.145)$ & & $U(2, 9)$ & $U(0, 2)$ & \\
    Sonora Elf Owl posteriors & & & & & & & & & \\
    classic & 4593 & $517^{+1}_{-2}$ & $<3.25$ & $-0.33\pm0.01$ & $0.317\pm0.005$ & & $3.99^{+0.02}_{-0.03}$ & $0.84\pm0.01$ & $-6.199\pm0.002$ \\
    GP & 0 & $549^{+2}_{-3}$ & $<3.25$ & $-0.30\pm0.02$ & $0.42\pm0.01$ & & $3.50\pm0.07$ & $0.81\pm0.01$ & $-6.190\pm0.005$ \\
    \hline
    ATMO2020++ (no PH$_3$) priors & & $U(250, 1200)$ & $U(2.5, 5.5)$ & $U(-1.0, 0.3)$ & & & & $U(0, 2)$ & \\
    ATMO2020++ (no PH$_3$) posteriors & & & & & & & & & \\
    classic & 4401 & $538\pm2$ & $4.94\pm0.02$ & $0.25\pm0.01$ & (0.55) & & ($5.12\pm0.04$) & $0.77\pm0.01$ & $-6.202\pm0.002$\\
    GP & 319 & $493\pm3$ & $4.14\pm0.02$ & $-0.09\pm0.02$ & (0.55) & & ($6.72\pm0.04$) & $1.01\pm0.01$ & $-6.186\pm0.002$ \\
    \hline
    ATMO2020++ priors & & $U(250, 1200)$ & $U(2.5, 5.5)$ & $U(-1.0, 0.3)$ & & & & $U(0, 2)$ & \\
    ATMO2020++ posteriors & & & & & & & & & \\
    classic & 4262 & $514\pm1$ & $4.80\pm0.01$ & $>0.30$ & (0.55) & & ($5.40\pm0.02$) & $0.93\pm0.01$ & $-6.179\pm0.002$ \\
    GP & 94 & $496^{+5}_{-3}$ & $4.30^{+0.04}_{-0.02}$ & $-0.02^{+0.03}_{-0.02}$ & (0.55) & & ($6.40^{+0.04}_{-0.08}$) & $1.03^{+0.01}_{-0.02}$ & $-6.163\pm0.002$\\
    \hline
    \end{tabular}
    \tablefoot{\small{\textit{First column:} log-Bayes factors relative to the Sonora Elf Owl (GP). \textit{Remaining columns:} Grid priors and posteriors. $U(a,b)$ refers to an uniform distribution between $a$ and $b$. The error bars correspond to the lower and upper bonds in the parameter space encompassing 68\% of the retrieved solutions around the best fit. Luminosities were computed using Equation (3) from \cite{zhang_disequilibrium_2024} to account for model–data discrepancies. Values inside parentheses are fixed by the model grids. In particular, the log(K$_\text{zz}$) values of the ATMO2020++ and ATMO2020++ (no PH$_3$) modeling results are linearly interpolated from the inferred log(g) using Figure 1 of \cite{phillips_new_2020}. }}
    \label{tab_parameters}
\end{table*}

\setlength{\tabcolsep}{7.5pt}
\renewcommand{\arraystretch}{1.5}
\begin{table*}[h!]
\tiny
    \centering
    \caption{\centering \small{GP hyperparameters priors and posteriors.}}
	\begin{tabular}{lcccccc}
    \hline
    \hline
    Parameter & log($a$)$_\text{FLAMINGOS-2}$ & log($a$)$_\text{NIRSpec}$ & log($a$)$_\text{MIRI}$ & log($l$)$_\text{FLAMINGOS-2}$ & log($l$)$_\text{NIRSpec}$ & log($l$)$_\text{MIRI}$\\
    Units & & & & log($\mu$m) & log($\mu$m) & log($\mu$m) \\
    \hline
    priors & $U(-0.5, 3)$ & $U(-0.5, 3)$ & $U(-0.5, 3)$ & $U(-4, 0)$ & $U(-4, 0)$ & $U(-4, 0)$ \\
    \hline
    BT-Settl posteriors & $0.01\pm0.05$ & $0.901\pm0.001$ & $2.23\pm0.02$ & $-1.86\pm0.03$ & $-2.691\pm0.001$ & $-1.701^{+0.001}_{-0.002}$ \\
    Sonora Diamondback posteriors & $0.33^{+0.02}_{-0.03}$ & $0.85\pm0.01$ & $1.62\pm0.02$ & $-1.65\pm0.01$ & $-2.725\pm0.001$ & $-1.676^{+0.003}_{-0.004}$ \\
    Sonora Elf Owl posteriors & $0.25^{+0.05}_{-0.06}$ & $0.60\pm0.01$ & $1.89\pm0.02$ & $-1.56^{+0.01}_{-0.06}$ & $-2.657^{+0.001}_{-0.002}$ & $-1.695^{+0.002}_{-0.003}$ \\
    ATMO2020++ (no PH$_3$) posteriors & $0.10^{+0.07}_{-0.06}$ & $0.75\pm0.01$ & $1.28\pm0.02$ & $-1.48^{+0.08}_{-0.06}$ & $-2.759\pm0.001$ & $-1.612\pm0.003$ \\
    ATMO2020++ posteriors & $0.09\pm0.07$ & $0.70\pm0.01$ & $1.24\pm0.02$ & $-1.51^{+0.08}_{-0.04}$ & $-2.753\pm0.001$ & $-1.606\pm0.002$ \\
    \hline
    \end{tabular}
    \label{tab_hyperparameters}
\end{table*}

\setlength{\tabcolsep}{18pt}
\renewcommand{\arraystretch}{1.5} 
\begin{table*}[h]
\tiny
    \centering
    \caption{\centering \small{Inversion results of Appendix \ref{app_inv}.}}
	\begin{tabular}{lcccccc}
    \hline
    \hline
    Parameter & $\ln \mathcal{B}$ & \Teff & log(g) & [M/H] & log($a$) & log($l$)\\
    Units & & (K) & (dex) & (dex) & & log($\mu$m) \\
    \hline
    priors & & $U(250, 1200)$ & $U(2.5, 5.5)$ & $U(-1.0, 0.3)$ & $U(-0.5, 2)$ & $U(-4, 0)$ \\
    posteriors classic & 229 & $471\pm1$ & $4.06\pm0.03$ & $-0.03\pm0.01$ & & \\
    posteriors GP & 0 & $476\pm4$ & $4.10^{+0.05}_{-0.06}$ & $-0.02\pm0.03$ & $0.01\pm0.05$ & $-1.97^{+0.03}_{-0.05}$ \\
    \hline
    injected & & 483 & 4.19 & 0.0 & 0.0 & -2.0 \\
    \hline
    \end{tabular}
    \label{tab_inj_test}
\end{table*}

\section{Models}

We used five atmospheric grids with different physical and chemical makeup : 

\begin{itemize}

\item BT-Settl \citep{allard_models_2012} is a grid that explores complex cloud microphysics and includes non-equilibrium chemistry as well as vertical mixing. Clouds are simulated by dividing the atmosphere into multiple layers and computing the distribution and size of grains by comparing their characteristic times of condensation, coalescence, dispersion and sedimentation.

\item Sonora Diamondback \citep{morley_sonora_2024} model grid includes vertical mixing and parameterized clouds. Clouds are parameterized using \cite{ackerman_precipitating_2001} approach with the sedimentation efficiency parameter f$_{sed}$. This model assumes radiative–convective and chemical equilibrium. We used a custom version of this grid extended at lower temperature by re-running the full forward model.

\item Sonora Elf Owl \citep{mukherjee_sonora_2024} is a cloudless grid that includes vertical mixing induced disequilibrium chemistry with sub-solar to super-solar [M/H] and C/O. The atmospheric models have been computed using the open-source radiative-convective equilibrium model \texttt{\textit{PICASO}}\footnote{\url{https://natashabatalha.github.io/picaso/}} \citep{Batalha2019, marley_sonora_2021, Mukherjee2023, JWST2023}. We used version v.2\footnote{\url{https://zenodo.org/records/15150865}} of the grid with a correction to the disequilibrium abundance of CO$_2$ and no PH$_3$ \citep{Wogan2025}.

\item ATMO2020++ and ATMO2020++ (no PH$_3$) \citep{phillips_new_2020, leggett_measuring_2021, meisner_exploring_2023} are extensions of the ATMO2020 model which additionally incorporate a non-adiabatic thermal structure of the atmospheres. These models are cloud-free but uses non-equilibrium chemical reactions of CO-CH$_4$ and N$_2$-NH$_3$, driving fingering convection in the atmosphere and changing the temperature gradient, to emulate their reddening effect.

\end{itemize}

\section{GP injection test}\label{app_inv}

This section describes the injection test we performed to probe the performance of the newly implemented GP. 

To ensure that the correlation amplitude and length scale retrieved during our inversion are consistent with the correlated noise present in real spectra, we constructed a synthetic Gemini/FLAMINGOS-2 spectrum. This mock spectrum was interpolated from the ATMO2020++ grid using atmospheric parameters of $T_{\rm eff} = 483$ K, $\log(g) = 4.19$, and [M/H] = 0.0 (see Table~\ref{table_sys}).

The model spectrum was convolved and sampled at a spectral resolution of $R_\lambda \sim 900$ to match that of FLAMINGOS-2, resulting in a noise-free spectrum $\vec{d}$. We then simulated observational noise using two components, as described by:

\begin{equation}
\vec{d_{\text{noisy}}} = \vec{d} + \vec{n_{\text{sky}}} + \vec{n_{\text{inst}}},
\end{equation}
where $\vec{n_{\text{sky}}}$ and $\vec{n_{\text{inst}}}$ represent the sky and instrumental noise contributions, respectively, and $\vec{d_{\text{noisy}}}$ is the final synthetic noisy spectrum.

We model the sky noise as uncorrelated Gaussian noise, assuming an average S/N of 10:

\begin{equation}
\vec{n_{\text{sky}}} \sim \mathcal{N}\left(0,\; \sigma^2\right), \quad \text{with} \quad \sigma^2 = \left(\frac{\bar{d}}{\mathrm{S/N}}\right)^2,
\end{equation}
where $\bar{d} = \mathrm{mean}(\vec{d})$. This component is the only one considered in the injected error bars.

To simulate the effect of correlated instrumental noise, we introduce a second component modeled with a GP characterized by a correlation length $\ell = 10^{-2}$~$\mu$m and amplitude $a = \sigma$ (i.e., strong correlated noise). This is represented by:

\begin{equation}
\vec{n_{\text{inst}}} \sim \mathcal{N}\left(0,\; \vec{C} \right), \quad \text{with} \quad C_{ij} = a^2 \exp\left[- \frac{\Delta \lambda_{i,j}^2}{2\ell^2} \right],
\end{equation}
where $\Delta \lambda_{i,j}$ denotes the wavelength separation between data points $i$ and $j$. Finally, we inverted this mock data with the ATMO2020++ using 1000 living points and uniform priors, with (in red) and without GP (in blue). Results are summarized in Fig. \ref{fig_inj_test} and Table \ref{tab_inj_test}. 

The top panel of Fig.~\ref{fig_inj_test} displays the fitted spectrum alongside the corner plots for both inversions. When spectral covariances are omitted, \texttt{\textit{ForMoSA}} fails to recover the injected values, and the posterior uncertainties are noticeably underestimated across all parameters. In contrast, incorporating a GP model improves performance: although some bias remains, the true values lie within the broader posterior distributions. Additionally, the GP-aided inversion successfully retrieves the covariance hyperparameters, yielding final estimates of $a = 1.0 \pm 0.1$ (true value: 1) and $\ell = 10.7^{+0.7}_{-0.1}$~nm (true value: 10~nm). The bottom panel of Fig.~\ref{fig_inj_test} presents the normalized residuals between the data and the model for the GP-aided inversion, clearly revealing correlated structures (in black). The red shaded contours represent the 1$\sigma$, 2$\sigma$, and 3$\sigma$ dispersions of an ensemble of 200 random draws from the final (fitted) covariance matrix $\vec{C}$. The inclusion of the GP kernel effectively captures both the structure and amplitude of the residuals.

\section{GP with additional atmospheric models}\label{app_othergrids}

In Sect.~\ref{sec3.2}, we focused our analysis on the two ATMO2020++ grids. Here, we provide an overview of the results obtained using the GP framework with the other grids (namely BT-Settl, Sonora Diamondback, and Sonora Elf Owl).

According to all statistical criteria adopted in this work, Sonora Elf Owl provides the preferred fit among the tested models (see Table~\ref{bayes_GP}). Nevertheless, the overall quality of the fit remains poor, particularly in its ability to reproduce the MIRI observations. In addition, the retrieved atmospheric parameters, most notably \Teff\ ($549^{+2}_{-3}$~K) and log(g) ($<3.25$~dex), remain inconsistent with the predictions of all evolutionary models explored in this study (see Table~\ref{evolutionary_tracks}).

On the other hand, the GP significantly improves the consistency of the retrieved \Teff and log(g) for BT-Settl and Sonora Diamondback, with \Teff~$=463\pm3$~K and log(g)~$=4.48^{+0.01}_{-0.02}$~dex for BT-Settl, and \Teff~$=515\pm1$~K and log(g)~$=4.500^{+0.003}_{-0.004}$~dex for Sonora Diamondback.

Similar to ATMO2020++, the metallicity is found to be sub-solar for all grids that explore this parameter, with [M/H]~$=-0.43\pm0.01$~dex and [M/H]~$=-0.30\pm0.02$~dex for Sonora Diamondback and Sonora Elf Owl, respectively. Contrary to the values retrieved with ATMO2020++ in this setup (see Sect.~\ref{sec4.1}), these metallicities are not consistent with the stellar metallicity of $0.00\pm0.08$~dex \citep{hojjatpanah_catalog_2019}. Similarly, the C/O ratio remains sub-solar, with C/O~$=0.42\pm0.01$ retrieved using Sonora Elf Owl.

The vertical mixing inferred with Sonora Elf Owl (\Kzz~=~$(3.2^{+0.6}_{-0.5})\times10^{3}$~cm$^2$\,s$^{-1}$) remains significantly lower than that predicted by the two ATMO2020++ grids (\Kzz~=~$(2.5^{+0.2}_{-0.4})\times10^{6}$~cm$^2$\,s$^{-1}$ and \Kzz~=~$(5.3^{+0.5}_{-0.5})\times10^{6}$~cm$^2$\,s$^{-1}$, with and without PH$_3$, respectively).

With the exception of the BT-Settl grid, the inferred GP correlation lengths are remarkably consistent across the different atmospheric models (see Table~\ref{tab_hyperparameters}). In contrast, we observe larger variations in the GP correlation amplitudes. This trend suggests that the correlation lengths may be primarily driven by each instrument's correlated noise pattern (see Sect.~\ref{sec4.2}), while the correlation amplitudes may instead reflect more the GP compensating for each specific model mismatches.

\end{appendix}

\end{document}